\documentclass[symmetry,article,accept,moreauthors,10pt,a4paper,pdftex]{mdpi}

\firstpage{1}
\articlenumber{x}
\doinum{10.3390/------}
\pubvolume{xx}
\pubyear{2016}
\copyrightyear{2016}
\externaleditor{Academic Editor: name}
\history{Received: date; Accepted: date; Published: date}
\pdfoutput=1

\usepackage{amsmath}
\usepackage{graphicx,epstopdf}
\usepackage{amssymb}
\usepackage{color}

\Title{A $\mathcal{PT}$-symmetric dual-core system with the sine-Gordon
nonlinearity and derivative coupling}

\Author{Jes\'us Cuevas--Maraver $^{1,2}$, Boris A. Malomed $^{3}$ and Panayotis G. Kevrekidis $^{4}$}
\AuthorNames{Jes\'us Cuevas--Maraver, Boris A. Malomed and Panayotis G. Kevrekidis}

\address{%
$^{1}$ \quad Grupo de F\'{\i}sica No Lineal, Departamento de F\'{\i}sica Aplicada I,
Universidad de Sevilla. Escuela Polit{\'e}cnica Superior, C/ Virgen de
\'Africa, 7, 41011-Sevilla, Spain; jcuevas@us.es\\
$^{2}$ \quad Instituto de Matem\'aticas de la Universidad de Sevilla (IMUS). Edificio
Celestino Mutis. Avda. Reina Mercedes s/n, 41012-Sevilla, Spain\\
$^{3}$ \quad Department of Physical Electronics, School of Electrical Engineering,
Faculty of Engineering, Tel Aviv University, Tel Aviv 69978, Israel; malomed@post.tau.ac.il\\
$^{4}$ \quad Department of Mathematics and Statistics, University of Massachusetts,
Amherst, MA 01003-4515, USA; kevrekid@math.umass.edu}

\corres{Correspondence: jcuevas@us.es; Tel.: +34-95-4556219}

\abstract{As an extension of the class of nonlinear $\mathcal{PT}$-symmetric models,
we propose a system of sine-Gordon equations, with the $\mathcal{PT}$
symmetry represented by balanced gain and loss in them. The equations are
coupled by sine-field terms and first-order derivatives. The sinusoidal
coupling stems from local interaction between adjacent particles in coupled
Frenkel-Kontorova (FK) chains, while the cross-derivative coupling, which
was not considered before, is induced by \emph{three-particle} interactions,
provided that the particles in the parallel FK\ chains move in different
directions. Nonlinear modes are then studied in this system. In particular,
kink-kink (KK) and kink-antikink (KA) complexes are explored by means of
analytical and numerical methods. It is predicted analytically and confirmed
numerically that the complexes are unstable for one sign of the sinusoidal
coupling, and stable for another. Stability regions are delineated in the
underlying parameter space. Unstable complexes split into free kinks and
antikinks, that may propagate or become quiescent, depending on whether they
are subject to gain or loss, respectively.}


\begin{document}

\section{Introduction}

Dual-core waveguides with intrinsic nonlinearity carried by each core offer
a convenient setting for the creation of stable dissipative solitons, by
application of linear gain to one core, and leaving the parallel-coupled
mate one lossy. This possibility was first proposed in the context of
nonlinear fiber optics in Ref. \cite{3}, see also a review in Ref. \cite{7}.
More recently, a similar scheme was elaborated for the application of gain
and stabilization of solitons in plasmonics \cite{8}, as well as for the
creation of stable two-dimensional dissipative solitons and solitary
vortices in dual laser cavities \cite{9}. Commonly adopted models of
dual-core nonlinear waveguides are based on linearly coupled systems of
nonlinear Schr\"{o}dinger (NLS)\ equations, which include gain and loss
terms \cite{7}. One of the advantages provided by these systems for the
theoretical analysis is the availability of exact analytical solutions for
stable solitons \cite{Atai}.

A crucial difference between dissipative solitons and their counterparts in
conservative media is the fact that the former ones exist as isolated
attractors, selected by the balance between gain and loss \cite{14}. On the
contrary, nonlinear conservative models, including those originating from
optics \cite{KA}, give rise to continuous families of soliton solutions,
rather than isolated ones.

A more special class of systems was identified at the interface between
conservative and dissipative ones, with spatially separated and precisely
balanced loss and gain. Such systems realize the $\mathcal{PT}$
(parity-time) symmetry, which originates in the quantum theory for
non-Hermitian Hamiltonians \cite{16}. A distinctive feature of the $\mathcal{%
PT}$-symmetric Hamiltonians is that they produce purely real spectra up to a
certain critical value of the strength of the part which represents the
balanced gain and loss. At the critical point, $\mathcal{PT}$-symmetry
breaking takes place, with the Hamiltonian's spectrum becoming complex above
this point.

Experimental implementation of the $\mathcal{PT}$ symmetry was suggested by
the fact that the propagation equation for optical beams in the paraxial
approximation has essentially the same form as the quantum-mechanical Schr%
\"{o}dinger equation, making it possible to emulate the evolution of the
wave function of a quantum particle by the transmission of an optical beam.
Accordingly, the implementation of the $\mathcal{PT}$ symmetry in optics was
proposed in Ref.\ \cite{18} and experimentally demonstrated in Ref.\ \cite%
{19} (see also review \cite{suchkov}), making use of mutually balanced
symmetrically placed gain and loss elements.

The presence of the Kerr effect in optical media suggests to consider the
interplay of the $\mathcal{PT}$ -symmetry with cubic nonlinearity, leading
to the prediction of $\mathcal{PT}$-symmetric solitons \cite{22}. A
crucially important issue in theoretical studies of such solitons is the
analysis of their stability, as the exact balance between the amplification
and dissipation may be easily disrupted \cite{23}. It was also proposed to
implement a similar setting in exciton-polariton condensates, where the gain
and pump are inherent ingredients of any setting~\cite{Nori}.

The above-mentioned couplers, with the gain and loss carried by different
parallel-coupled cores, offer a natural platform for the realization of the $%
\mathcal{PT}$ symmetry in optics and other physical settings, if the gain
and loss in the two cores are exactly balanced \cite{coupler,coupler2}.
Adding the intrinsic Kerr nonlinearity, the analysis makes it possible to
find $\mathcal{PT}$-symmetric solitons in the coupler and their stability
boundary in an exact analytical form \cite{coupler}. In addition to the
fundamental two-component solitons, higher-order ones \cite{Radik} and
soliton chains \cite{LuLi} in the $\mathcal{PT}$-symmetric coupler were also
considered. Bilayer systems of other types with balanced gain and loss were
investigated too \cite{bilayer}.

A relevant extension of the analysis is to combine the linear $\mathcal{PT}$
symmetry with other physically relevant nonlinearities, such as the
sine-Gordon (SG) one \cite{SG-PT}, or the second-harmonic-generating
quadratic nonlinearity, which can be readily implemented in optics too \cite%
{quadratic}. The SG nonlinearity finds its realizations in a broad range of
physical settings \cite{book}, including various forms of the
Frenkel-Kontorova (FK)\ model \cite{Kosevich,FK}, long Josephson junctions
(JJs) between bulk superconductors \cite{JJ}, self-induced transparency \cite%
{SIT-sG}, ferromagnets \cite{Ivanov}, ferroelectrics \cite{Maugin}, and
field-theory models \cite{field}. In all these realizations, fundamental
dynamical modes are topological solitons (kinks and antikinks) \cite{kivs89}%
, including fluxons and antifluxons, i.e., magnetic-flux quanta trapped in
long JJs \cite{JJ}. The consideration of systems combining the SG
nonlinearity with the $\mathcal{PT}$ symmetry is relevant, in particular,
because it suggests a possibility to establish a link between the
phenomenology underlain by the $\mathcal{PT}$ symmetry in optics with
similar phenomena in other physical settings.

Following this direction, a natural possibility is to implement the $%
\mathcal{PT}$ symmetry in couplers, composed of an amplified core and a
dissipative one, as outlined above, in the case when each core carries the
SG dynamics. Previously, many works have addressed models based on coupled
SG equations \cite{coupled-SG,kivs89}, such as those modeling double FK
chains \cite{Kosevich}, stacked JJs \cite{stacked-JJ,Nori2,Stan} and the
layered structure of high-temperature superconductors \cite{highTc}.
However, the competition of the gain and loss acting in the two coupled SG
cores was not considered before. This is the subject of the present work.
The basic $\mathcal{PT}$-symmetric coupled-SG\ model is formulated in Sec.
II, where conditions for the background stability of the system are derived
too. The dual system is supported by two couplings, one presented by
previously known sinusoidal terms \cite{Kosevich,Stan}, and a previously
unexplored coupling, based on the first-order cross-derivatives, which
represent a \emph{three-body} interaction between adjacent particles in an
underlying double FK chain, with different directions of motion of particles
in the two individual chains. Fundamental topological modes in the system
are built as kink-kink (KK) and kink-antikink (KA) complexes (in addition to
them, nontopological small-amplitude breathers are briefly considered too).
Analytical and numerical results for the KK and KA states are reported in
Secs. III and IV. The most essential results are existence and stability
boundaries for the KK and KA states, delineated in the underlying parameter
space by means of analytical and numerical methods. The paper is concluded
by Sec. V, suggesting also some potential extensions to future work.

\section{The model}

\subsection{The coupled sine-Gordon system}

The $\mathcal{PT}$-symmetric system of coupled SG equations for a real
amplified field, $\phi \left( x,t\right) $, and an attenuated one, $\psi
\left( x,t\right) $, is adopted in the following form:

\begin{eqnarray}
\phi _{tt}-\phi _{xx}+\sin \phi &=&\epsilon \sin \left( \phi -\psi \right)
+\beta \psi _{x}+\alpha \phi _{t},  \label{phi} \\
\psi _{tt}-\psi _{xx}+\sin \psi &=&\epsilon \sin \left( \psi -\phi \right)
-\beta \phi _{x}-\alpha \psi _{t},  \label{psi}
\end{eqnarray}%
where coefficient $\alpha $ represents the balanced gain and loss, and $%
\epsilon $ is the coefficient of the inter-chain sinusoidal coupling in the
double FK chain \cite{Kosevich} (a similar coupling appears in a triangular
system of three long JJs with a trapped magnetic flux \cite{Stan}). Such
sinusoidal coupling has also been considered extensively in terms of the
so-called sine-lattices~\cite{takhomma1}, which represent, e.g.,
base-rotator models of the DNA double helix~\cite{takhomma2}. Coefficient $%
\beta $ in Eqs. (\ref{phi}) and (\ref{psi}) represents a new type of the
anti-symmetric cross-derivative coupling between the two SG equations (note
that reflection $x\rightarrow -x$ makes it possible to fix $\beta >0$). It
is different from the usual magnetic coupling between stacked JJs, which
would be represented by symmetric second-order cross-derivatives \cite%
{stacked-JJ}. The derivation of this coupling in terms of coupled FK chains
is outlined below.

The Hamiltonian corresponding to the conservative version of Eqs. (\ref{phi}%
) and (\ref{psi}), with $\alpha =0$, is

\begin{gather}
H=\int_{-\infty }^{+\infty }\left[ \frac{1}{2}\left( \phi _{t}^{2}+\psi
_{t}^{2}+\phi _{x}^{2}+\psi _{x}^{2}\right) +\left( 1-\cos \phi \right)
+\left( 1-\cos \psi \right) \right.  \notag \\
\left. -\epsilon \left( 1-\cos \left( \phi -\psi \right) \right) -\beta \phi
\psi _{x}\right] dx  \label{H}
\end{gather}%
(the last term in the integrand may be replaced by its symmetrized form, $%
-\left( \beta /2\right) \left( \phi \psi _{x}-\psi \phi _{x}\right) $). The $%
\mathcal{PT}$ transformation for the system of Eqs. (\ref{phi}) and (\ref%
{psi}) is defined as follows:
\begin{equation}
\phi =\tilde{\psi},\psi =\tilde{\phi},x=-\tilde{x},t=-\tilde{t}.
\label{tilde}
\end{equation}%
It includes the swap of $\psi $ and $\phi $ as the $\mathcal{P}$
transformation in the direction transverse to $x$, as in usual $\mathcal{PT}$%
-symmetric couplers \cite{coupler,coupler2}. Obviously, the system is
invariant with respect to transformation (\ref{tilde}).

\begin{figure}[tbp]
\includegraphics[width=15cm]{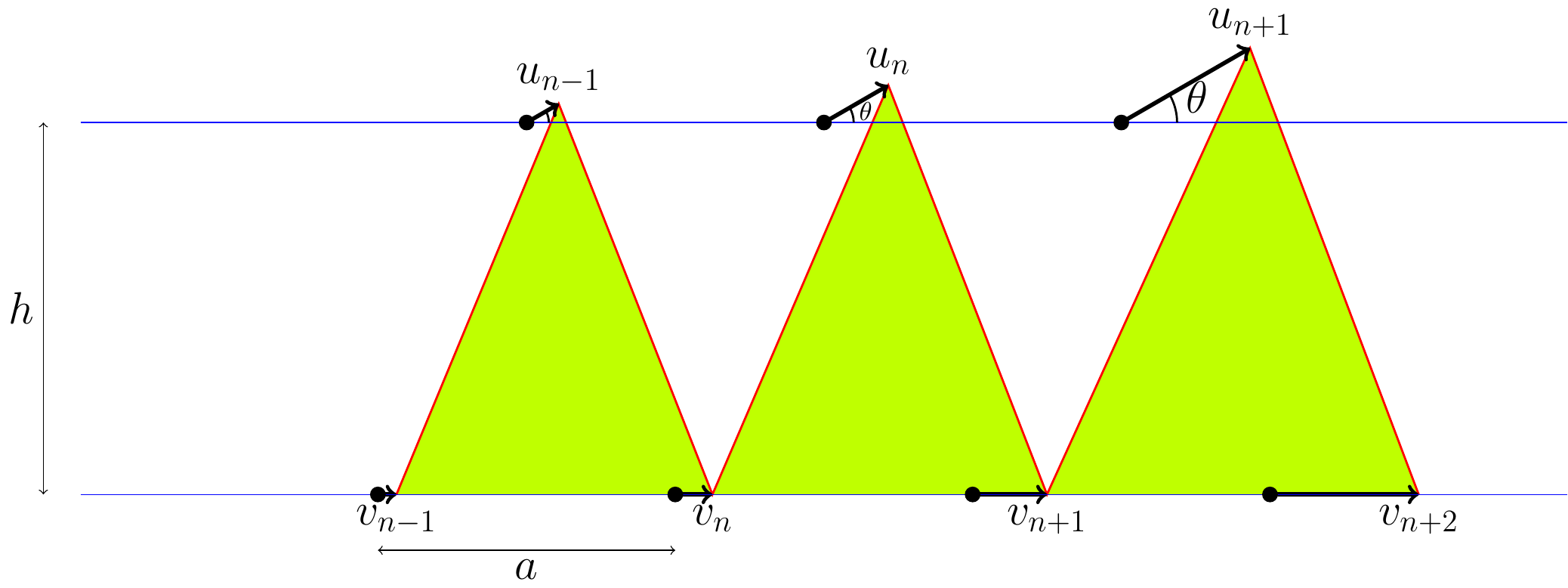}
\caption{(Color online) Three cells of the underlying double FK chain.
Vectors show displacements of the particles in the top and bottom chains.
The energy of the local two-particle coupling between the chains, which
gives rise to terms $\sim \protect\epsilon $ in Eqs. (\protect\ref{phi}) and
(\protect\ref{psi}), is determined by squared lengths of the red links, see
Eq. (\protect\ref{ll}). The total energy of the three-particle coupling,
which generates terms $\sim \protect\beta $ in Eqs. (\protect\ref{phi}) and (%
\protect\ref{psi}), is proportional to the combined area of the shaded
triangles, see Eq. (\protect\ref{area}). The polarization angle $\protect%
\theta $ of the motion in the top chain determines the relative strength of
the two couplings in Eqs. (\protect\ref{phi}) and (\protect\ref{psi}) as per
Eq. (\protect\ref{theta}). Other details of the setting are explained in the
text.}
\label{FK_sketch}
\end{figure}

The anti-symmetric cross-derivative coupling emerges in a \textquotedblleft
triangular" dual FK system schematically displayed in Fig. \ref{FK_sketch},
where $a$ and $h$ are the spacing in each FK chain, and the separation
between the parallel chains, respectively. It is assumed that particles with
coordinates $v_{n}(t)$ belonging to the bottom chain may move in the
horizontal direction, while particles with coordinates $u_{n}(t)$, which
belong to the top chain, move along a different direction, under fixed angle
$\theta $ with respect to the horizontal axis. The inner energies of the two
chains (the interaction between them is considered below) can be written as%
\begin{gather}
E_{\mathrm{inner}}=\sum_{n}\left\{ \frac{m}{2}\left[ \left( \frac{du_{n}}{dt}%
\right) ^{2}+\left( \frac{dv_{n}}{dt}\right) ^{2}\right] \right.  \notag \\
+\frac{\kappa }{2}\left[ \left( u_{n}-u_{n-1}\right) ^{2}+\left(
v_{n}-v_{n-1}\right) ^{2}+2a(\cos \theta )\left( u_{n}-u_{n-1}\right)
+2a\left( v_{n}-v_{n-1}\right) +2a^{2}\right]  \notag \\
\left. W_{0}\left[ \left( 1-\cos \left( \frac{2\pi u_{n}}{b}\right) \right)
+\left( 1-\cos \left( \frac{2\pi v_{n}}{b}\right) \right) \right] \right\} ,
\label{E}
\end{gather}%
where $m$ is the mass of each particle, $\kappa $ is the strength of the
elastic coupling along each chain (equal coefficients in front of $\left(
u_{n}-u_{n-1}\right) ^{2}$ and $\left( v_{n}-v_{n-1}\right) ^{2}$ is a
consequence of identity $\cos ^{2}\theta +\sin ^{2}\theta \equiv 1$), $W_{0}$
is the depth on the onsite potential, and $b$ is its period. The model may
have $b\ll a$, if the FK chains are built as superlattices on top of an
underlying lattice potential; then, kinks, which are considered below,
represent a relatively weak deformation of the chains. The equations of
motion are generated by the energy as usual,%
\begin{equation}
m\frac{d^{2}\left\{ u_{n},v_{n}\right\} }{dt^{2}}=-\frac{\partial E}{%
\partial \left\{ u_{n},v_{n}\right\} }.  \label{motion}
\end{equation}%
In the continuum approximation, which corresponds to

\begin{equation}
\frac{2\pi }{b}\left\{ u_{n}(t),v_{n}(t)\right\} \rightarrow \left\{ \phi
\left( x,t\right) ,\psi \left( x,t\right) \right\} ,~an\rightarrow x,
\label{cont-limit}
\end{equation}%
inner energy (\ref{E}) generates the corresponding terms in Eqs. (\ref{phi})
and (\ref{psi}), while terms $\sim 2a(\cos \theta )\left(
u_{n}-u_{n-1}\right) $ and $2a\left( v_{n}-v_{n-1}\right) $ in Eq. (\ref{E})
carry over into derivatives $\phi _{x}$ and $\psi _{x}$, which give no
contribution into the dynamical equations.

Proceeding to the consideration of the coupling between the top and bottom
chains, we note that the usual local coupling may be interpreted as produced
by energies of diagonal springs linking adjacent particles. These energies
are, in turn, proportional to squared lengths of these links. In particular,
the sum of the squared lengths for the pair of links connecting the $n$-th
particle in the top chain to its neighbors in the bottom chain, with numbers
$n$ and $n+1$, is

\begin{gather}
l_{n,n}^{2}+l_{n,n+1}^{2}=2\left( h+u_{n}\sin \theta \right) ^{2}+\left(
\frac{a}{2}+u_{n}\cos \theta -v_{n}\right) ^{2}+\left( \frac{a}{2}-u_{n}\cos
\theta +v_{n+1}\right) ^{2}  \notag \\
\equiv \frac{a^{2}}{2}+2h^{2}+a\left( v_{n+1}-v_{n}\right) +4h\left( \sin
\theta \right) u_{n}+2u_{n}^{2}+v_{n}^{2}+v_{n+1}^{2}-2\left( \cos \theta
\right) u_{n}\left( v_{n}+v_{n+1}\right) ,  \label{ll}
\end{gather}
A straightforward consideration of the continuum limit (\ref{cont-limit})
for the corresponding FK\ Hamiltonian, demonstrates that the last term in
expression (\ref{ll}) represents the local-coupling energy, which indeed
gives rise to the linearized form [$\sin \left( \phi -\psi \right) \approx
\phi -\psi $]\ of terms $\sim \epsilon $ in\ Eqs. (\ref{phi}) and (\ref{psi}%
), so that $\cos \theta \sim -\epsilon $ (the most essential results are
obtained below for $\epsilon <0$, which thus corresponds to $\cos \theta >0$%
). Other terms in Eq. (\ref{ll}) do not represent the inter-chain coupling,
and may be absorbed into an appropriate definition of the Hamiltonian of
each chain. In particular, term $a\left( v_{n+1}-v_{n}\right) $ becomes a
derivative $\sim \psi _{x}$ in the continuum limit, hence this term drops
out from the continuum Hamiltonian, as mentioned above.

On the other hand, the cross-derivative couplings $\sim \beta $ in Eqs. (\ref%
{phi}) and (\ref{psi}) may be induced by \emph{three-particle} interactions
(TPIs) in the underlying FK system, while usual binary interactions cannot
give rise to this coupling (FK\ models with TPIs were considered in few
previous works \cite{TPI}). In the simplest case, the energy of the relevant
TPI can be defined to be proportional to the sum of areas of the respective
three-body triangles, as shown in Fig. \ref{FK_sketch}. (i.e., the TPI is
induced by the \textquotedblleft surface tension" of the triangles).\
Interactions of this type may be realized in heterogeneous structures with
FK chains attached to nanolayers which provide the surface tension, such as
graphene (by dint of a technique similar to that reported in Ref. \cite%
{stack-graphene}) or other materials (see, e.g., Ref. \cite{n=0}). This type
of the TPI may also be adopted as a relatively simple phenomenological model.

To derive the cross-derivative-coupling term in the Hamiltonian induced by
the TPI, we note that the area of the triangle, confined by the links whose
lengths are given by Eq. (\ref{ll}), is

\begin{equation}
A_{n,n,n+1}=\frac{1}{2}\left[ ah+h\left( v_{n+1}-v_{n}\right) +\left( \sin
\theta \right) au_{n}+\left( \sin \theta \right) \left( v_{n+1}-v_{n}\right)
u_{n}\right] .  \label{area}
\end{equation}%
The transition to the continuum limit as per Eq. (\ref{cont-limit})
transforms the last term in Eq. (\ref{area}) into the last term in the
Hamiltonian density corresponding to Eq. (\ref{H}), with $\beta \sim -\sin
\theta $, while term $h\left( v_{n+1}-v_{n}\right) $ in expression (\ref%
{area}), similar to its above-mentioned counterpart $a\left(
v_{n+1}-v_{n}\right) $ in Eq. (\ref{ll}), becomes a full derivative, $\psi
_{x}$, in the continuum limit, hence it may be dropped from the Hamiltonian.
Also, term $\left( \sin \theta \right) au_{n}$ may be absorbed into the
definition of the inner-chain FK Hamiltonian.

Lastly, the above considerations demonstrate that the ``polarization
angle'', $\theta $, in the dual-FK chain (see Fig. \ref{FK_sketch})
determines the relative strength of the two couplings:

\begin{equation}
\beta /\epsilon \sim \tan \theta .  \label{theta}
\end{equation}%
This relation shows that the cross-derivative coupling, $\beta \neq 0$,
emerges when the motion directions in the coupled FK chains are not parallel
($\theta \neq 0$), while the usual coupling, with $\epsilon \neq 0$, acts
unless the two directions are mutually perpendicular, $\theta \neq \pi /2$.

\subsection{Conditions for stability of the flat states}

As the main objective of the work is to produce KK and KA complexes, which
interpolate between flat states with $\phi ,\psi =0~(\mathrm{mod}~2\pi )$,
and explore stability of the complexes, a necessary preliminary condition is
the stability of the flat states against small perturbations. To address
this issue, we use the linearized version of Eqs. (\ref{phi}) and (\ref{psi}%
), which governs the evolution of small perturbations added to the flat
states:

\begin{eqnarray}
\phi _{tt}-\phi _{xx}+\left( 1-\epsilon \right) \phi +\epsilon \psi &=&\beta
\psi _{x}+\alpha \phi _{t},  \label{lin-phi} \\
\psi _{tt}-\psi _{xx}+\left( 1-\epsilon \right) \psi +\epsilon \phi
&=&-\beta \phi _{x}-\alpha \psi _{t}.  \label{lin-psi}
\end{eqnarray}%
The substitution of the usual ansatz for eigenmodes of small perturbations, $%
\left\{ \phi ,\psi \right\} =\left\{ \phi _{0},\psi _{0}\right\} \exp \left(
ikx-i\omega t\right) ,$ in Eqs. (\ref{lin-phi}) and (\ref{lin-psi}) yields
the biquadratic dispersion equation for $\omega $:

\begin{equation}
\left( \omega ^{2}\right) ^{2}-2\left( 1-\epsilon +k^{2}-\frac{\alpha ^{2}}{2%
}\right) \omega ^{2}+\left( 1-\epsilon +k^{2}\right) ^{2}-\epsilon
^{2}-\beta ^{2}k^{2}=0.  \label{quadr}
\end{equation}%
Straightforward algebraic manipulations demonstrate that Eq. (\ref{quadr})
gives rise to a purely real, i.e., instability-free spectrum, with $\omega
^{2}(k^{2})\geq 0$ for all $k^{2}\geq 0$, under the following conditions:

\begin{eqnarray}
\epsilon &\leq &1/2,  \label{1} \\
\beta ^{2} &\leq &2\left[ \left( 1-\epsilon \right) +\sqrt{1-2\epsilon }%
\right] ,  \label{2} \\
\alpha ^{2} &\leq &\beta ^{2},  \label{3} \\
\alpha ^{2} &\leq &2\left[ \left( 1-\epsilon \right) -\sqrt{1-2\epsilon }%
\right] .  \label{4}
\end{eqnarray}%
Note that, if $\epsilon $ satisfies constraint (\ref{1}), then the
expression on the right-hand side of Eq. (\ref{4}) is always non-negative,
i.e., the corresponding stability interval for $\alpha ^{2}$ exists.

\subsection{The small-amplitude limit: coupled NLS equations}

Small-amplitude solutions to Eqs. (\ref{phi}) and (\ref{psi}) for
oscillatory nontopological solitons (breathers) may be looked for as

\begin{equation}
\phi \left( x,t\right) =2e^{-i\left( 1-\epsilon /2\right) t}U(x,t)+\mathrm{%
c.c.},~\psi \left( x,t\right) =2e^{-i\left( 1-\epsilon /2\right) t}V(x,t)+%
\mathrm{c.c.},  \label{uv}
\end{equation}%
where $U$ and $V$ are small-amplitude slowly varying complex functions, and $%
\mathrm{c.c.}$ stands for the complex conjugate. In the lowest nontrivial
approximation, the complex amplitudes obey a system of coupled NLS
equations, which are derived from Eqs. (\ref{phi}) and (\ref{psi}) by the
substitution of ansatz (\ref{uv}):

\begin{eqnarray}
iU_{t}+\frac{1}{2}U_{xx}+|U|^{2}U &=&\frac{\epsilon }{2}V-\frac{1}{2}\beta
V_{x}+\frac{1}{2}i\alpha U,  \label{u} \\
iV_{t}+\frac{1}{2}V_{xx}+|V|^{2}V &=&\frac{\epsilon }{2}U+\frac{1}{2}\beta
U_{x}-\frac{1}{2}i\alpha V.  \label{v}
\end{eqnarray}%
The dispersion relation for the linearized version of Eqs. (\ref{u}) and (%
\ref{v}) yields

\begin{equation}
\omega =\frac{1}{2}\left( k^{2}\pm \sqrt{\epsilon ^{2}-\alpha ^{2}+\beta
^{2}k^{2}}\right) ,  \label{omega}
\end{equation}%
cf. Eq. (\ref{quadr}), hence the zero-background solution is stable under
the condition of
\begin{equation}
\alpha ^{2}\leq \epsilon ^{2},  \label{<}
\end{equation}
which is not affected by coefficient $\beta $, cf. Eqs. (\ref{1})-(\ref{4}).
Note that the expansion of Eq. (\ref{4}) for small $\epsilon $ leads to the
same condition, illustrating the consistency of the analysis.

The system of coupled NLS equations (\ref{u}) and (\ref{v}) with $\beta =0$
is identical to the above-mentioned model of the $\mathcal{PT}$-symmetric
coupler, which may be realized in terms of nonlinear fiber optics, admitting
an exact analytical solution for solitons and their stability \cite{coupler}%
. The additional terms $\sim \beta $ in the coupler model may represent
the temporal dispersion of the coupling strength in fiber optics \cite%
{Chiang} (in that case, $\ t$ and $x$ are replaced, respectively, by the
propagation distance and reduced time \cite{KA}). Solitons and their
stability in the framework of Eqs. (\ref{u}) and (\ref{v}) with $\beta \neq
0 $ constitute a separate problem which will be considered elsewhere. Here,
we mention that a broad class of exact solutions to Eqs. (\ref{u}), (\ref{v}%
) can be found in the case of \textit{supersymmetry} \cite{super}, $\alpha
=\epsilon $ (note that it is precisely the edge of the stability region (\ref%
{<})), when both the gain and loss coefficients in the two cores are exactly
equal to the coefficient of the coupling between them. Indeed, in this case,
the substitution of%
\begin{equation}
\left\{ U\left( x,t\right) ,V\left( x,t\right) \right\} =\left\{
1,-i\right\} \tilde{U}\left( x.t\right) \exp \left( \frac{i}{2}\beta x+\frac{%
i}{8}\beta ^{2}t\right)  \label{UV}
\end{equation}%
reduces Eqs. (\ref{u}), (\ref{v}) to a single integrable NLS equation,%
\begin{equation}
i\tilde{U}_{t}+\frac{1}{2}\tilde{U}_{xx}+|\tilde{U}|^{2}\tilde{U}=0.
\label{NLS}
\end{equation}%
Equation (\ref{NLS}) gives rise to the commonly known vast set of single-
and multisoliton solutions \cite{KA}, which generate the respective
two-component solutions via Eq. (\ref{UV}). The analysis of the stability of
this solutions is a subject for a separate work.

\section{Analytical results for kink-kink (KK)\ and kink-antikink (KA)\
complexes}

\subsection{Stationary equations}

Quiescent solutions to Eqs. (\ref{phi}) and (\ref{psi}), $\phi (x)$ and $%
\psi (x)$, satisfy the stationary equations,

\begin{eqnarray}
\frac{d^{2}\phi }{dx^{2}} &=&\sin \phi -\epsilon \sin \left( \phi -\psi
\right) -\beta \frac{d\psi }{dx},  \label{stat-phi} \\
\frac{d^{2}\psi }{dx^{2}} &=&\sin \psi -\epsilon \sin \left( \psi -\phi
\right) +\beta \frac{d\phi }{dx},  \label{stat-psi}
\end{eqnarray}%
which may be considered (with $x$ formally replaced by time) as equations of
two-dimensional motion of a mechanical particle of unit mass in the plane
with coordinates $\left( \phi ,\psi \right) $, under the action of the
Lorentz force $\sim \beta $ and a force produced by an effective potential, $%
U\left( \phi ,\psi \right) =-\left( 1-\cos \phi \right) -\left( 1-\cos \psi
\right) +\epsilon \left[ 1-\cos \left( \phi -\psi \right) \right] ,$ cf.
expression (\ref{H}) for the Hamiltonian of the SG system. As indicated
above, we are interested in solutions for KK and KA complexes interpolating
between different flat states, i.e., fixed points (FPs) of Eqs. (\ref%
{stat-phi}) and (\ref{stat-psi}), $\phi _{0}$ and $\psi _{0}$. The FPs are
determined by equations

\begin{gather}
\sin \phi _{0}=-\sin \psi _{0},  \notag \\
\epsilon \sin \left( \phi _{0}-\psi _{0}\right) =\sin \phi _{0}.
\label{equil}
\end{gather}%
Equation (\ref{equil}) gives rise to three sets of the FPs,

\begin{eqnarray}
\phi _{0} &=&\psi _{0}=2\pi n,  \label{2n2n} \\
\phi _{0} &=&\psi _{0}=\pi \left( 1+2n\right) ,  \label{1+2n 1+2n} \\
\phi _{0} &=&2\pi n,~\psi _{0}=\pi \left( 2n\pm 1\right) ,  \label{2n 1+2n}
\\
\phi _{0} &=&\pi \left( 2n\pm 1\right) ,~\psi _{0}=2\pi n,  \label{1+2n 2n}
\end{eqnarray}%
with arbitrary integer $n$. In addition to that, at $|\epsilon |>1/2$ there also exist FPs with%
\begin{equation}
\left( \phi _{0},\psi _{0}\right) =\pm \arccos \left( \frac{1}{2\epsilon }%
\right) +2\pi n.  \label{>}
\end{equation}

It is easy to see that the Hamiltonian density defined by Eq. (\ref{H}) has
a minimum only at FP (\ref{2n2n}), while FPs (\ref{1+2n 1+2n}) and (\ref{2n
1+2n}), (\ref{1+2n 2n}) correspond to a maximum or saddle points,
respectively, hence stable FPs may be produced solely by Eq. (\ref{2n2n})
[for this reason, the detailed stability analysis, which produces conditions
(\ref{1})-(\ref{4}), was presented above only for this type of the FP]. The
KK and KA complexes should connect the FPs with different values of $n$,
hence these complexes represent heteroclinic trajectories of the dynamical
system based on Eqs. (\ref{stat-phi}) and (\ref{stat-psi}).

As concerns FP (\ref{>}), it is straightforward to check that, at $\epsilon
<-1/2$, they correspond to an absolute maximum of the Hamiltonian density,
therefore they are unstable. On the other hand, at $\epsilon >1/2$, when FP (%
\ref{2n2n}) is unstable, according to Eq. (\ref{1}), FP (\ref{>}) realizes
an absolute minimum of the Hamiltonian density, hence it may represent a
\emph{stable} flat solution. Heteroclinic solutions connecting such FPs can
be constructed, but they are different from $2\pi $ kinks. In particular,
for $0<2\epsilon -1\ll 1$ and $\beta =0$, an approximate solutions
connecting two FPs (\ref{>}) with $n=0$ and opposite signs chosen for $\pm $
is%
\begin{equation}
\phi (x)=-\psi (x)\approx \sqrt{2\left( 2\epsilon -1\right) }\tanh \left(
\sqrt{\left( 2\epsilon -1\right) x/2}\right) .  \label{tanh}
\end{equation}%
Detailed consideration of such heteroclinic solutions at $\epsilon >1/2$ is
beyond the scope of the present work.

The gain-loss coefficient, $\alpha $, does not appear in Eqs. (\ref{stat-phi}%
) and (\ref{stat-psi}), therefore it has no bearing on the shape of
stationary states. Nevertheless, $\alpha $ does affect stability of KK and
KA complexes, as shown below. This is similar to what has been shown earlier
in $\mathcal{PT}$-symmetric SG models in~Ref. \cite{SG-PT}.

\subsection{Exact KK and KA solutions for $\protect\beta =0$}

In the case of $\beta =0$, stationary equations (\ref{stat-phi}) and (\ref%
{stat-psi}) admit two obvious types of solutions. One of them is symmetric,

\begin{equation}
\phi _{0}(x)=\psi _{0}(x),  \label{=+}
\end{equation}%
with $\phi _{0}(x)$ being any stationary solution of the usual sine-Gordon
equation,

\begin{equation}
\frac{d^{2}\phi _{0}}{dx^{2}}=\sin \phi _{0},  \label{sin}
\end{equation}%
such as the $2\pi $ kink, antikink, or periodic kink chains.

In the absence of the gain and loss terms, $\alpha =0$, the stability of the
symmetric solutions with $\beta =0$ can be investigated in the general form.
Indeed, eigenmodes of small perturbations added to solution (\ref{=+}) can
be looked for as symmetric or antisymmetric ones:

\begin{equation}
\left\{ \phi (x,t),\psi \left( x,t\right) \right\} =\left\{ \phi
_{0}(x),\phi _{0}(x)\right\} +\zeta e^{-i\omega _{\epsilon }t}\phi
_{1}^{\left( \pm \right) }(x)\left\{ 1,\pm 1\right\} ,  \label{+-}
\end{equation}%
where $\zeta $ is an infinitesimal amplitude of the perturbation, $\omega
_{\epsilon }$ is an eigenfrequency of the perturbation mode, and $\phi
_{1}^{(\pm )}(x)$ are the eigenmodes themselves. The stability condition is
that all eigenfrequencies must be real, $\omega _{\epsilon }^{2}\geq 0$.

The linearization of the nonstationary coupled SG equations (\ref{phi}) and (%
\ref{psi}) (with $\beta =\alpha =0$) leads to the following equations for
the modal functions, $\phi _{1}^{\left( \pm \right) }(x)$:

\begin{eqnarray}
\omega _{\epsilon }^{2}\phi _{1}^{(+)} &=&-\frac{d^{2}\phi _{1}^{(+)}}{dx^{2}%
}+\left[ \cos \phi _{0}(x)\right] \phi _{1}^{(+)},  \label{+} \\
\left( \omega _{\epsilon }^{2}+2\epsilon \right) \phi _{1}^{(-)} &=&-\frac{%
d^{2}\phi _{1}^{(-)}}{dx^{2}}+\left[ \cos \phi _{0}(x)\right] \phi
_{1}^{(-)}.  \label{-}
\end{eqnarray}%
Obviously, solutions of Eq. (\ref{+}) have the same eigenvalues, $\omega
_{\epsilon }^{2}\equiv \omega _{0}^{2}$, as in the case of the usual single
SG equation. In particular, the usual SG $2\pi $ kink (or antikink), see Eq.
(\ref{kink}) below, is commonly known to be stable, hence it gives rise to $%
\omega _{0}^{2}\geq 0$. Thus, the symmetric perturbations cannot destabilize
the KK complex in the coupled system.

On the other hand, Eq. (\ref{-}) for the antisymmetric perturbations gives
rise to eigenvalues

\begin{equation}
\omega _{\epsilon }^{2}=\omega _{0}^{2}-2\epsilon .  \label{eps}
\end{equation}%

Note that Eq. (\ref{+}) has a \textit{zero-mode} solution, with $\omega
_{0}=0$:

\begin{equation}
\phi _{1}^{(0)}(x)=\frac{d\phi _{0}}{dx}  \label{zero}
\end{equation}%
(because $d\phi _{0}/dx$ for the $2\pi $ kink has no zeros at finite $x$,
the fact that eigenmode (\ref{zero}) corresponds to $\omega _{0}^{2}=0$
confirms, via the Sturm theorem \cite{Sturm}, that all higher-order
eigenmodes generated by Eq. (\ref{+}), that have zero crossings, correspond
to $\omega _{\epsilon }^{2}>0$, which corroborates the stability of the KK
complex against the symmetric perturbations). Then, the substitution of $%
\omega _{0}^{2}=0$ in Eq. (\ref{eps}) gives rise to $\omega _{\epsilon
}^{2}=-2\epsilon $. Thus, the symmetric KK solution of the coupled SG system
is \emph{unstable }for $\epsilon >0$, and \emph{stable} for $\epsilon <0$.

The other type of solutions to stationary equations (\ref{stat-phi}) and (%
\ref{stat-psi}) with $\beta =0$ is antisymmetric,

\begin{equation}
\phi _{0}(x)=-\psi _{0}(x),  \label{=-}
\end{equation}%
with $\phi _{0}(x)$ being any solution of the stationary\textit{\ double SG}
equation,

\begin{equation}
\frac{d^{2}\phi _{0}}{dx^{2}}=\sin \phi _{0}-\epsilon \sin \left( 2\phi
_{0}\right) .  \label{dsin}
\end{equation}%
An exact $2\pi $-kink solution to Eq. (\ref{dsin}), which corresponds to the
antisymmetric KA complex produced by the coupled SG system, has the known
form \cite{Bullough}:

\begin{equation}
\phi _{0}(x)=\pi +2\arctan \left( \frac{\sinh \left( \sqrt{1-2\epsilon }%
x\right) }{\sqrt{1-2\epsilon }}\right) .  \label{exact}
\end{equation}%
In the limit of $\epsilon =0$, this solution is tantamount to the usual $%
2\pi $ kink. Obviously, solution (\ref{exact}) exists at $\epsilon <1/2$,
which agrees with condition (\ref{1}). In the limit of $\epsilon =1/2$, Eq. (%
\ref{exact}) takes a limit form, which is a relevant solution too, in this
special case:

\begin{equation}
\phi _{0}^{(\epsilon =1/2)}(x)=\pi +2\arctan x.  \label{1/2}
\end{equation}

At $\epsilon >1/2$, a valid solution can be obtained from Eq. (\ref{exact})
by analytic continuation, in the form of a spatially periodic state
(essentially, a KA chain):

\begin{equation}
\phi _{0}(x)=\pi +2\arctan \left( \frac{\sin \left( \sqrt{2\epsilon -1}%
x\right) }{\sqrt{2\epsilon -1}}\right) .  \label{continued}
\end{equation}

All KA chains are unstable \cite{book}, hence solution (\ref{continued}) is
unstable too.

The analysis of the stability of the KA complex, represented by solution (%
\ref{exact}) in the framework of coupled SG equations (\ref{phi}) and (\ref%
{psi}), is not analytically tractable, even in the case of $\beta =\alpha =0$%
. The respective numerical results are presented below -- see, in
particular, Figs. \ref{fig:stab1}, \ref{fig:stab2} and \ref{fig:stab3}.

\subsection{Perturbative solutions for small $\protect\beta $}

If the cross-derivative coupling constant $\beta $ in Eqs. (\ref{stat-phi})
and (\ref{stat-psi}) is treated as a small perturbation, it is easy to see
that the first correction to the symmetric KK\ solution (\ref{=+}), which
was obtained above for $\beta =0$, is antisymmetric. Thus, the full
(approximate) solution becomes asymmetric at $\beta \neq 0$:

\begin{equation}
\left\{ \phi (x),\psi (x)\right\} =\left\{ \phi _{0}(x)+\beta \phi
_{1}(x),~\phi _{0}(x)-\beta \phi _{1}(x)\right\} ,  \label{anti}
\end{equation}%
with perturbation $\phi _{1}(x)$ determined by the linearized equation

\begin{equation}
\frac{d^{2}\phi _{1}}{dx^{2}}-\left( \cos \phi _{0}(x)\right) \phi
_{1}+2\epsilon \phi _{1}=-\frac{d\phi _{0}}{dx}.  \label{epsilon}
\end{equation}%
An \emph{exact solution} to Eq. (\ref{epsilon}) can be found, making use of
the zero mode (\ref{zero}):

\begin{equation}
\phi _{1}(x)=-\frac{1}{2\epsilon }\frac{d\phi _{0}}{dx}.  \label{phi1}
\end{equation}

In particular, the unperturbed $2\pi $ kink/antikink is

\begin{equation}
\phi _{0}(x)=4\arctan (e^{\sigma x}),  \label{kink}
\end{equation}%
with polarity $\sigma =+1/-1$. The respective perturbed KK solution (\ref%
{anti}) is

\begin{equation}
\left\{ \phi (x),\psi (x)\right\} =\left\{ 4\arctan (e^{\sigma x})-\frac{%
\beta \sigma }{\epsilon \cosh x},~4\arctan (e^{\sigma x})+\frac{\beta \sigma
}{\epsilon \cosh x}\right\} .  \label{phipsi}
\end{equation}%
Below, these approximate analytical results are compared to their
numerically found counterparts in Fig. \ref{fig:profiles}.

Similarly, the first correction to the antisymmetric KA solution (\ref{=-}),
induced by small $\beta $, must be symmetric, cf. Eq. (\ref{anti}), hence
the corresponding approximate solution is again asymmetric:

\begin{equation*}
\left\{ \phi (x),\psi (x)\right\} =\left\{ \phi _{0}(x)+\beta \phi
_{1}(x),-\phi _{0}(x)+\beta \phi _{1}(x)\right\} ,
\end{equation*}%
with perturbation $\phi _{1}$ determined by the linearized equation (cf. Eq.
(\ref{epsilon}))

\begin{equation}
\frac{d^{2}\phi _{1}}{dx^{2}}-\left( \cos \phi _{0}(x)\right) \phi _{1}=%
\frac{d\phi _{0}}{dx}.  \label{no-epsilon}
\end{equation}%
However, it is not obvious how to identify a solution to Eq. (\ref%
{no-epsilon}) in an exact form, unlike solution (\ref{phi1}) of Eq. (\ref%
{epsilon}). Below, stationary solutions for the KA complexes are found in a
numerical form, see Fig. \ref{fig:profiles}.

\section{Numerical results for kink-kink (KK) and kink-antikink (KA)
complexes}

\subsection{Stationary KK and KA solutions and stability equations}

In this section, we report results for KK and KA solutions of the full
coupled system of SG equations (\ref{phi}) and (\ref{psi}) and their
stability, obtained by means of numerical methods and complementing the
analytical results of the previous section. Computations were performed with
the help of a finite-difference scheme for the spatial derivatives, using a
central-difference scheme for the first-order ones, and free-end (Neumann)
boundary conditions.

\begin{figure}[tbp]
\begin{tabular}{cc}
\includegraphics[width=7cm]{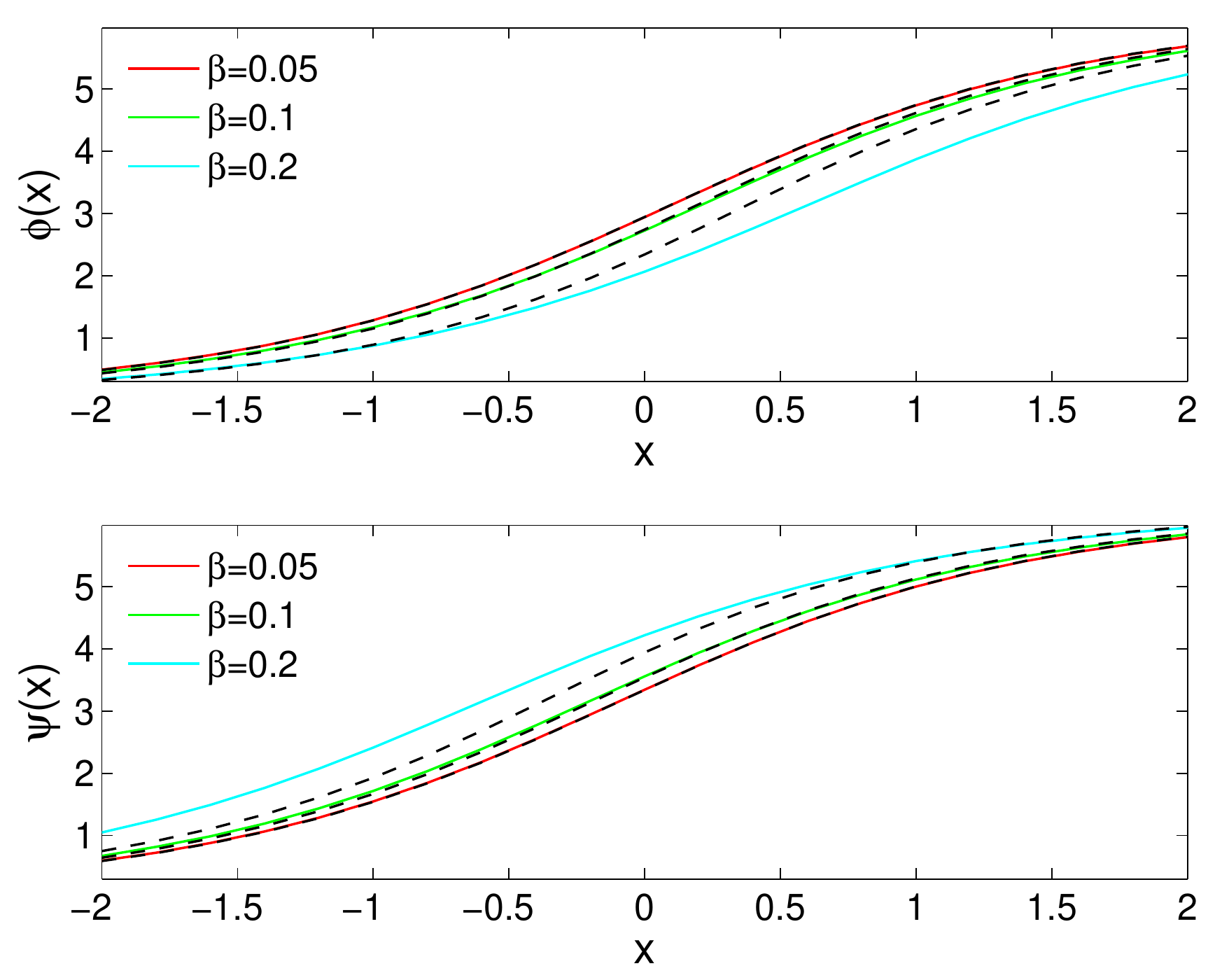} & %
\includegraphics[width=7cm]{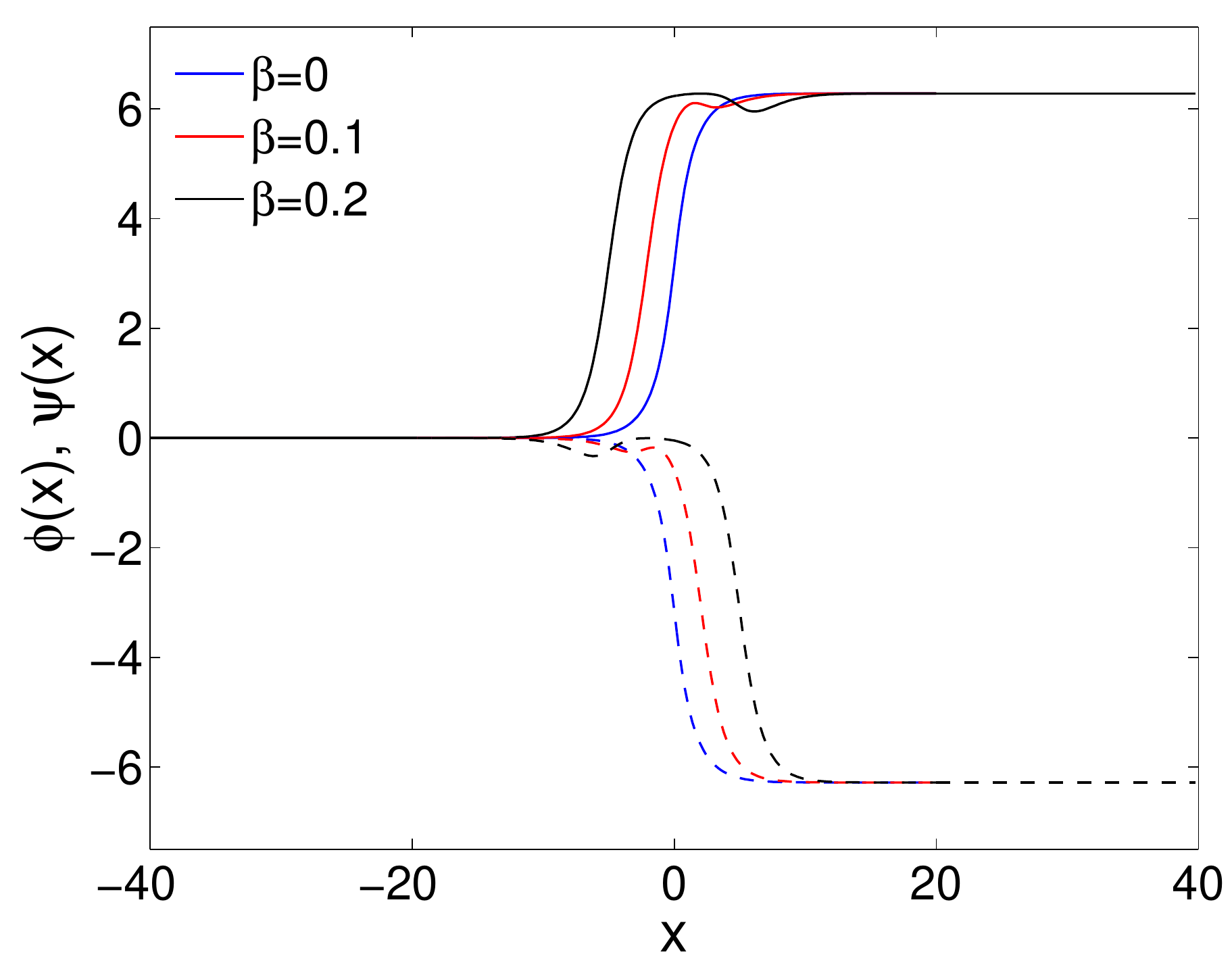} \\
\includegraphics[width=7cm]{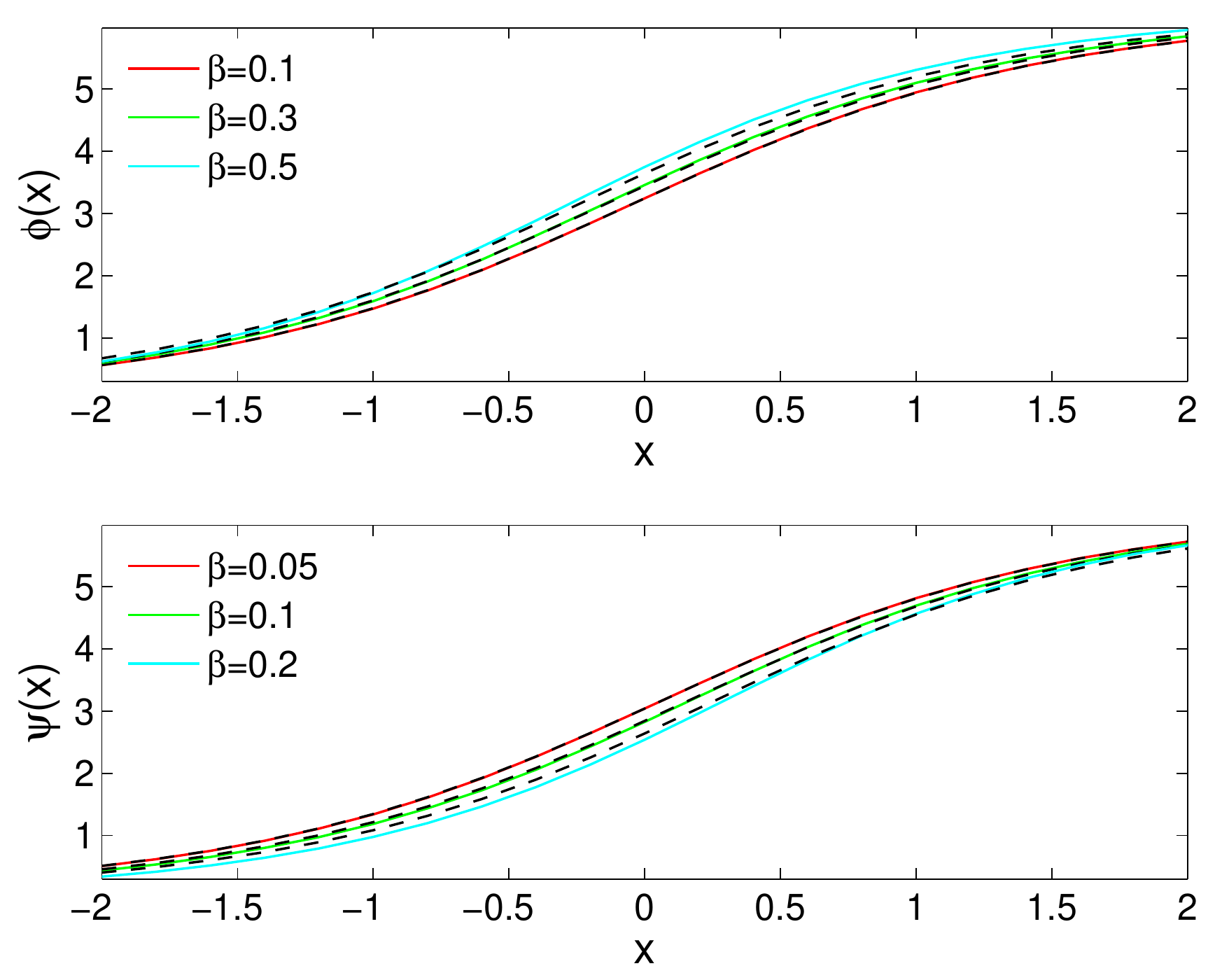} & %
\includegraphics[width=7cm]{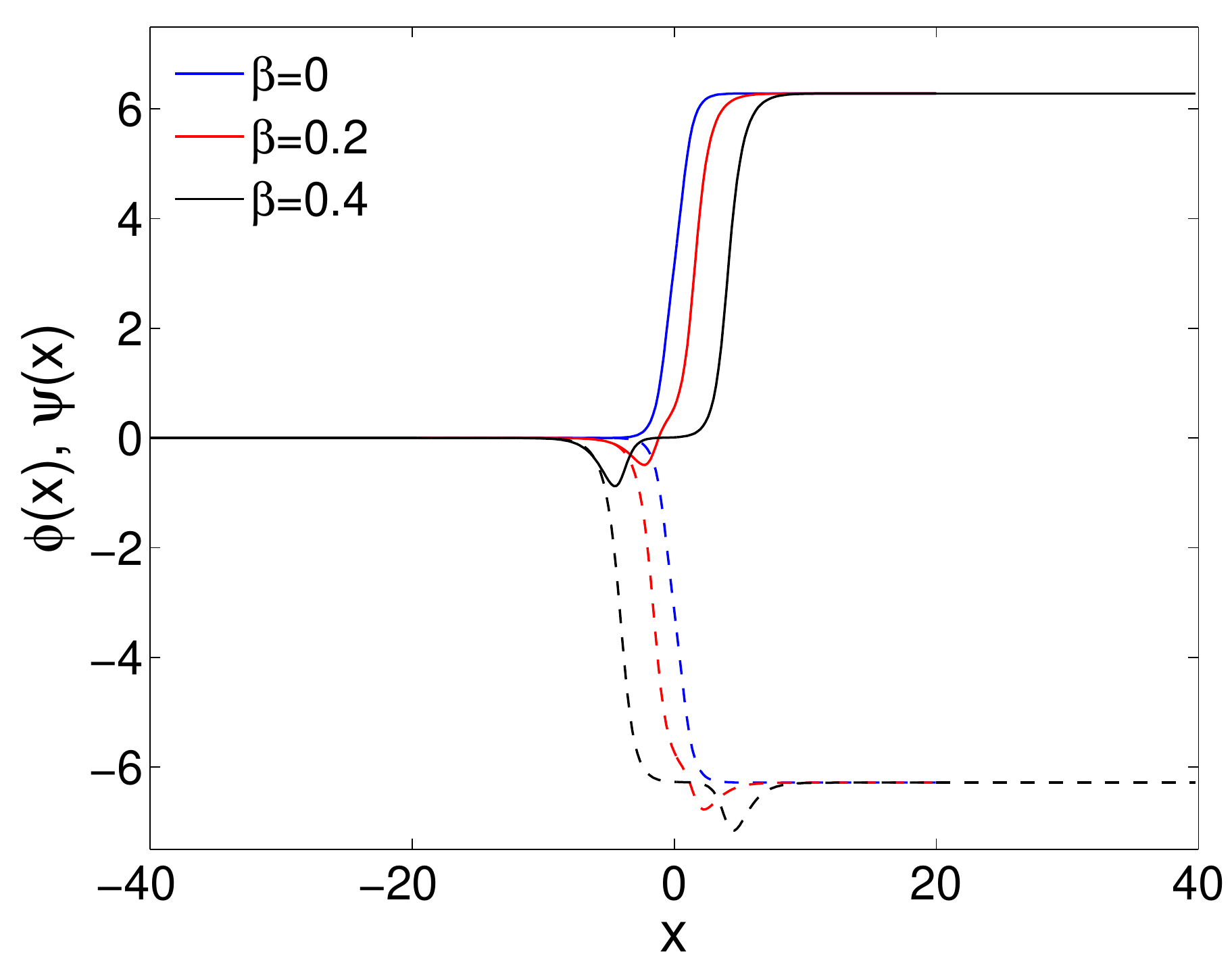}%
\end{tabular}%
\caption{(Color online) Profiles of the KK (left) and KA (right) complexes
for $\protect\epsilon =0.25$ (top) and $\protect\epsilon =-1$ (bottom), in
both cases. In the former case, only the central parts of $\protect\phi (x)$
and $\protect\psi (x)$ are shown, and the profiles are compared to the
approximate analytical solution given by Eq.~(\protect\ref{phipsi}), which
is depicted by dashed lines. In the latter case, both $\protect\phi (x)$ and
$\protect\psi (x)$ are shown in the same plot, respectively by full and
dashed lines, respectively. The solutions are unstable at $\protect\epsilon %
>0$ (in the top panels) and stable at $\protect\epsilon \leq 0$ (in the
bottom panels). }
\label{fig:profiles}
\end{figure}

Figure \ref{fig:profiles} shows profiles of solutions of both the KK and KA
types at $\epsilon >0$ and $\epsilon <0$. For the KK complexes, we display
the central part of the $\phi $ and $\psi $ components, and compare them to
the perturbative prediction given by Eq. (\ref{phipsi}); for the KA modes,
we display both components only in the numerical form, as an analytical
solution of perturbative equation~(\ref{no-epsilon}) is not available.

To study the spectral stability of the solutions, we proceed by adding small
perturbations to the stationary solutions as follows:

\begin{gather}
\phi (x,t)=\phi _{0}(x)+\delta u_{1}(x)\mathrm{e}^{\lambda t},  \notag \\
\psi (x,t)=\psi _{0}(x)+\delta u_{2}(x)\mathrm{e}^{\lambda t},  \notag \\
\phi _{t}(x,t)=\delta v_{1}(x)\mathrm{e}^{\lambda t},  \notag \\
\psi _{t}(x,t)=\delta v_{2}(x)\mathrm{e}^{\lambda t},  \label{pert}
\end{gather}%
where $\lambda $ is a (generally, complex) (in)stability eigenvalue, $%
u_{1,2}(x)$ and $v_{1,2}(x)$ being the corresponding eigenmodes of the small
perturbations. The spectral stability condition is that there should not
exist eigenvalues with Re$(\lambda )>0$. The substitution of expression (\ref%
{pert}) into Eqs. (\ref{phi})-(\ref{psi}) and the subsequent linearization
to $\mathcal{O}(\delta )$ leads to the following eigenvalue problem:

\begin{equation}
\lambda \left(
\begin{array}{c}
u_{1} \\
u_{2} \\
v_{1} \\
v_{2}%
\end{array}%
\right) =\left(
\begin{array}{cccc}
0 & 0 & I & 0 \\
0 & 0 & 0 & I \\
\epsilon \cos (\phi _{0}-\psi _{0})-\cos (\phi _{0})+\partial _{xx} &
-\epsilon \cos (\phi _{0}-\psi _{0})+\beta \partial _{x} & \alpha I & 0 \\
-\epsilon \cos (\phi _{0}-\psi _{0})-\beta \partial _{x} & \epsilon \cos
(\phi _{0}-\psi _{0})-\cos (\psi _{0})+\partial _{xx} & 0 & -\alpha I%
\end{array}%
\right) \left(
\begin{array}{c}
u_{1} \\
u_{2} \\
v_{1} \\
v_{2}%
\end{array}%
\right) ,  \label{eq:stability}
\end{equation}%
where $I$ is the identity operator. As said above, the KK and KA profiles do not
depend on the gain-loss coefficient $\alpha $, although their stability does
depend on $\alpha $, through its explicit inclusion inside the matrix of
Eq.~(\ref{eq:stability}).

Instabilities are not only caused by the existence of localized KK or KA
complexes, but can also emerge from the background if conditions (\ref{1})-(%
\ref{4}) are not fulfilled. In particular, if only Eqs. (\ref{1}) and/or (%
\ref{2}) are violated, background eigenvalues with nonzero real parts
possess a zero imaginary part. On the other hand, if solely Eqs. (\ref{3})
and/or (\ref{4}) are violated, the respective unstable background
eigenvalues have nonzero imaginary parts, with a slight difference between
the two cases: if condition (\ref{3}) is violated, the eigenvalues with
nonzero real parts are those which possess the largest imaginary part,
whereas, if condition (\ref{4}) fails, unstable eigenvalues arise with the
smallest imaginary part (i.e., close to wavenumbers $k=0$).

\subsection{Instability of the KK and KA complexes at $\protect\epsilon >0$}

\begin{figure}[tbp]
\begin{tabular}{cc}
\includegraphics[width=7cm]{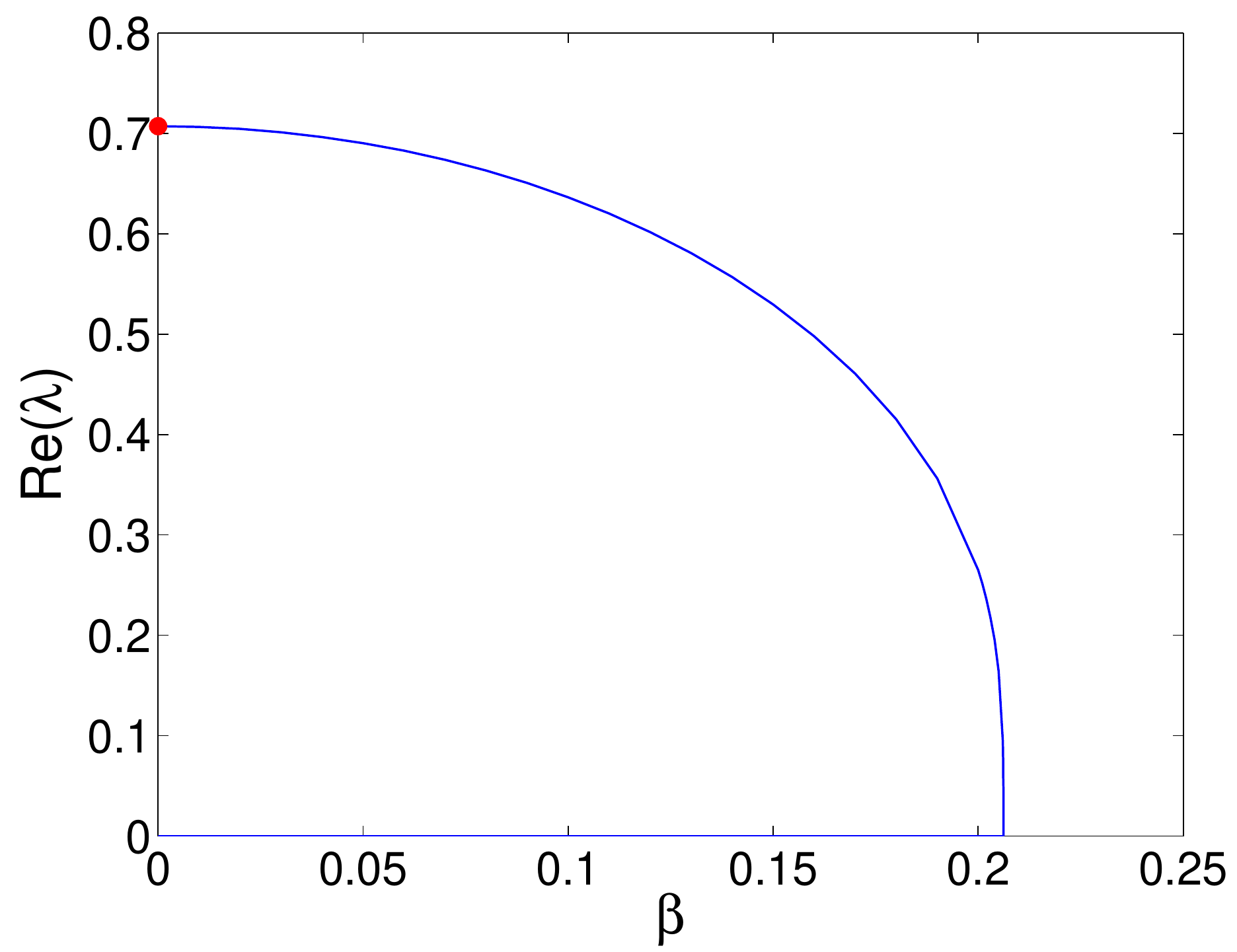} & %
\includegraphics[width=7cm]{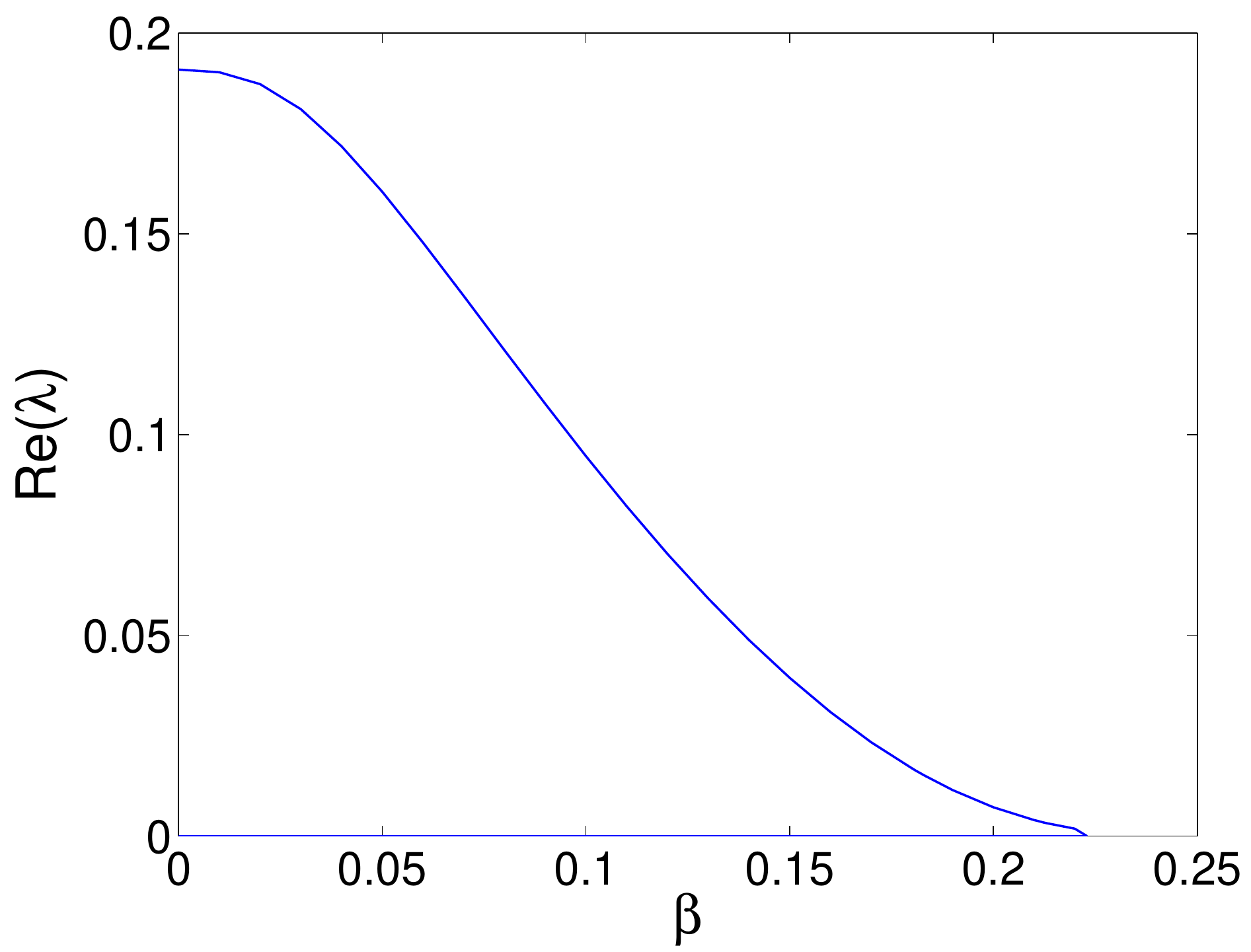}%
\end{tabular}%
\caption{(Color online) The real part of the instability eigenvalues versus $%
\protect\beta $ for unstable KK (left) and KA (right) complexes at $\protect%
\alpha =0$ and $\protect\epsilon =0.25$. They exist up to the point at which
Re$(\protect\lambda )$ vanishes. The red dot at the left panel shows the
exact prediction for the zero mode of (\protect\ref{eps}).}
\label{fig:stab1}
\end{figure}

\begin{figure}[tbp]
\includegraphics[width=7cm]{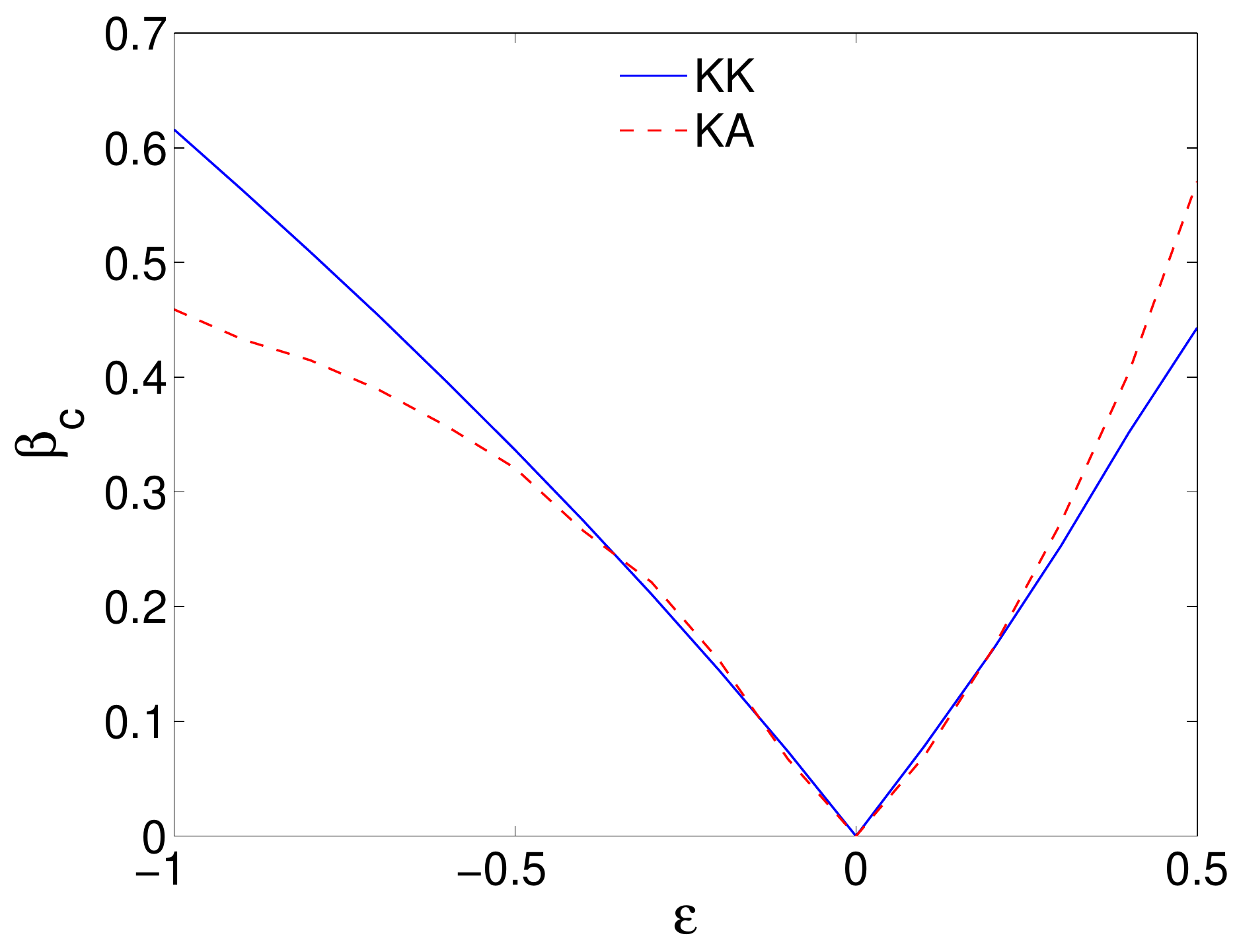}
\caption{(Color online) The critical value of the cross-derivative coupling,
$\protect\beta _{c}$, above which the KK and KA complexes do not exist,
versus $\protect\epsilon $. Note that both complexes exist beyond the right
edge (at $\protect\epsilon \geq 0.5$); however, as condition (\protect\ref{1}%
) does not hold in that area, the flat states are unstable in it. {The
picture is independent of $\protect\alpha $, as only stability and dynamical
properties depend of this parameter.}}
\label{fig:plane1}
\end{figure}

Both KK and KA complexes are found to be exponentially unstable at $\epsilon
>0$, in agreement with the exact analytical result obtained above for the KK
complexes in the case of $\beta =\alpha =0$ [see Eq. (\ref{eps})]. Further,
the complexes of both types exist, at $\epsilon >0$, below a critical point,
$\beta <\beta _{c}$, which depends on $\epsilon $. For given $\epsilon $,
the critical values $\beta _{c}$ are different for the KK and KA solutions.
At $\beta =\beta _{c}$, the real eigenvalue responsible for the kink's
instability vanishes. Actually, $\beta _{c}$ satisfies inequality (\ref{2}),
i.e., the existence range is smaller than the range of the background
stability whenever Eqs. (\ref{3}) and (\ref{4}) hold. Figure \ref{fig:stab1}
shows the dependence of the real part of the instability eigenvalues on $%
\beta $ for fixed $\epsilon =0.25$. Figure \ref{fig:plane1} shows the
dependence of $\beta_{c}$ on $\epsilon$ for both KK and KA complexes.

\begin{figure}[tbp]
\begin{tabular}{cc}
\includegraphics[width=7cm,bb=0 0 576 432,clip=on]{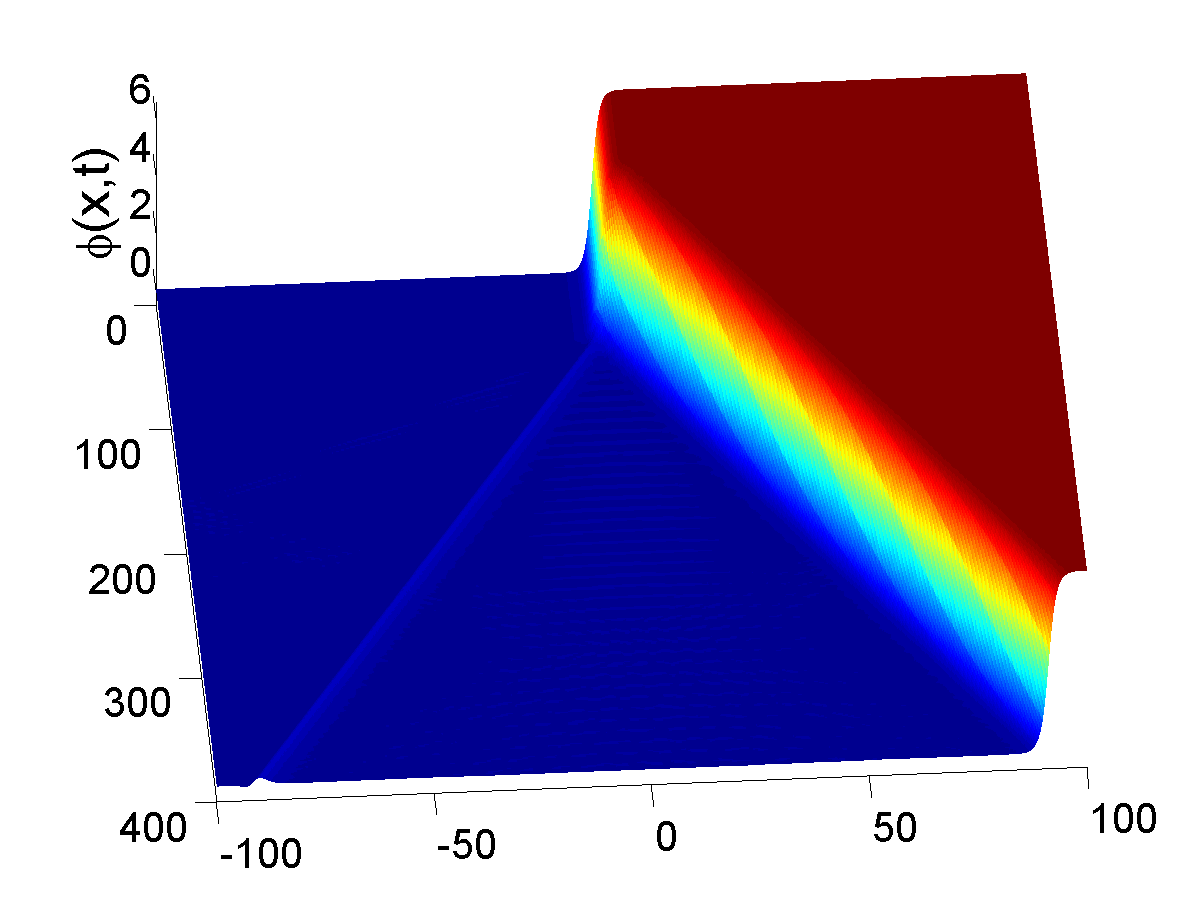} & %
\includegraphics[width=7cm,bb=0 0 576 432,clip=on]{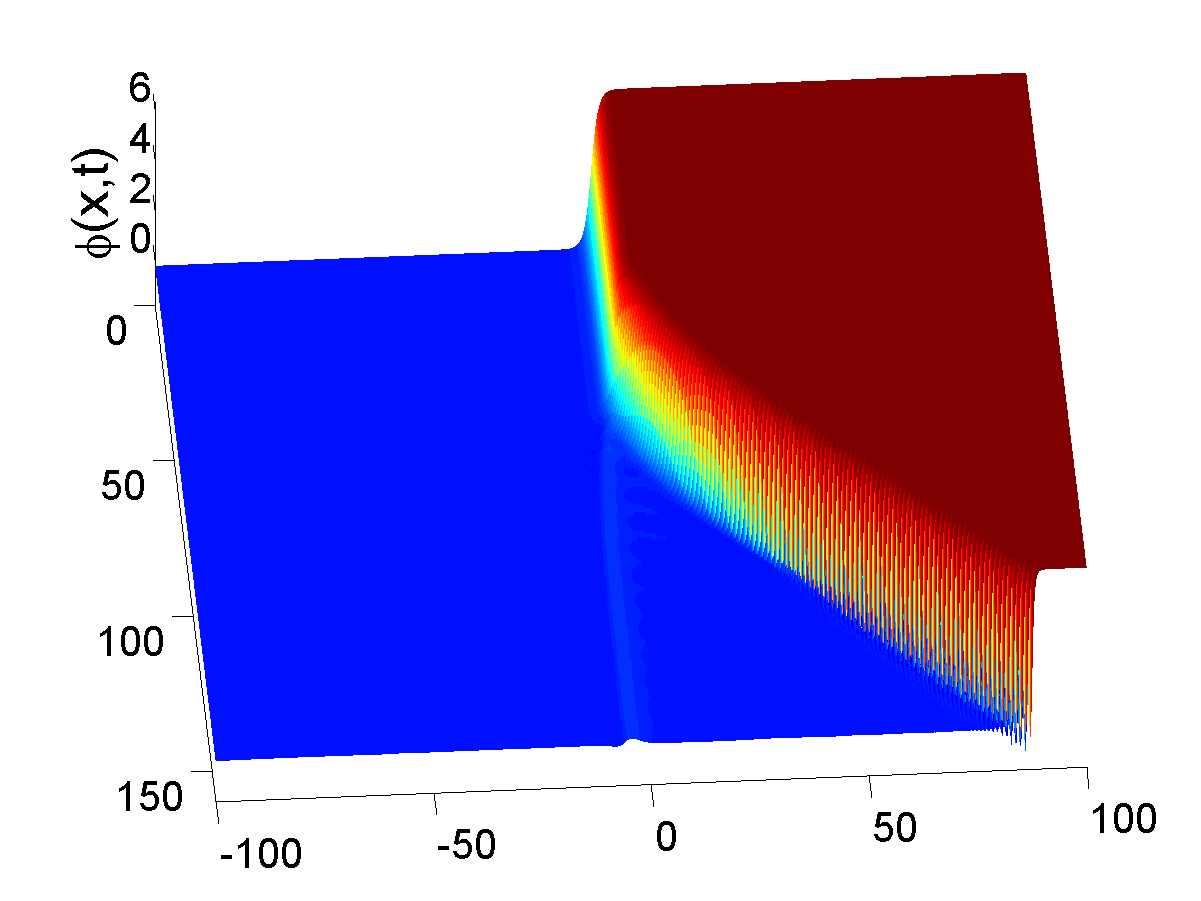} \\
\includegraphics[width=7cm,bb=0 0 576 432,clip=on]{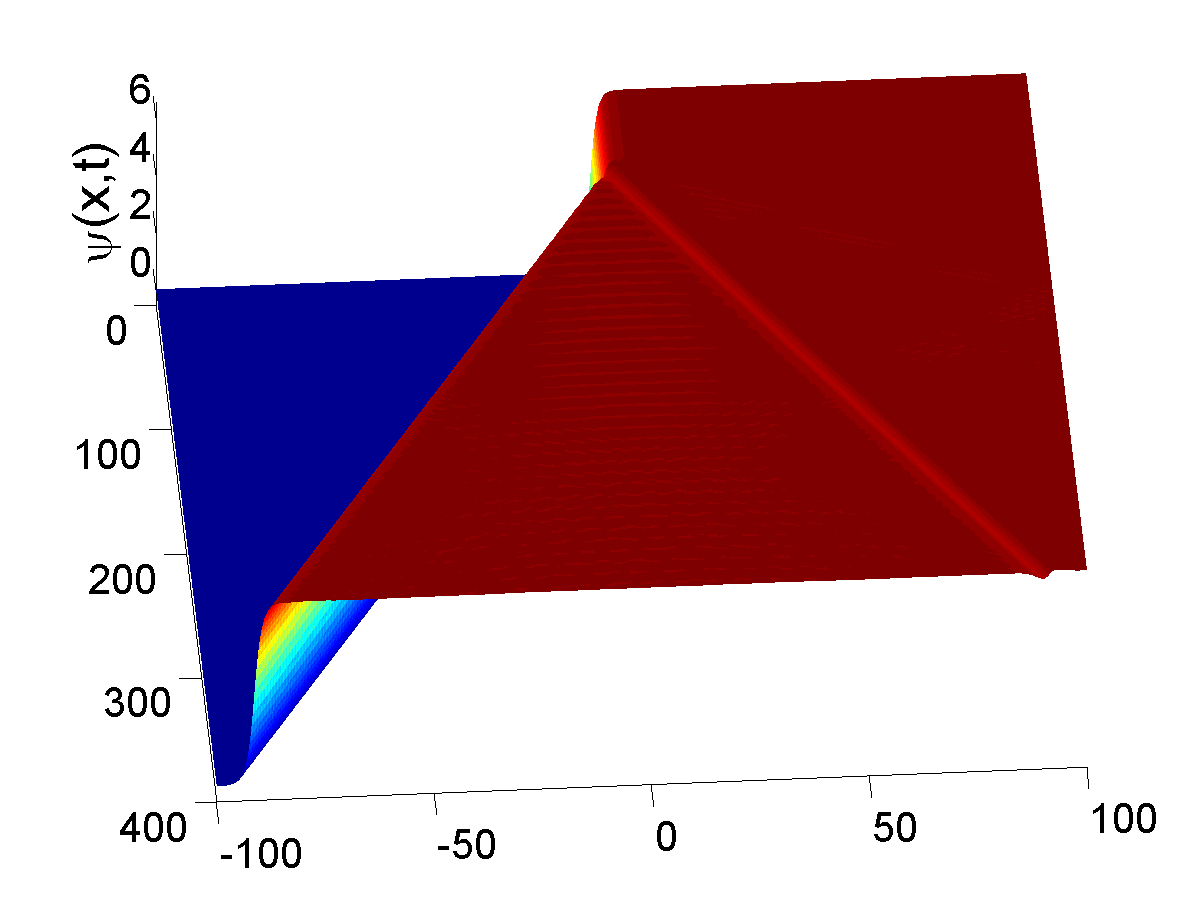} & %
\includegraphics[width=7cm,bb=0 0 576 432,clip=on]{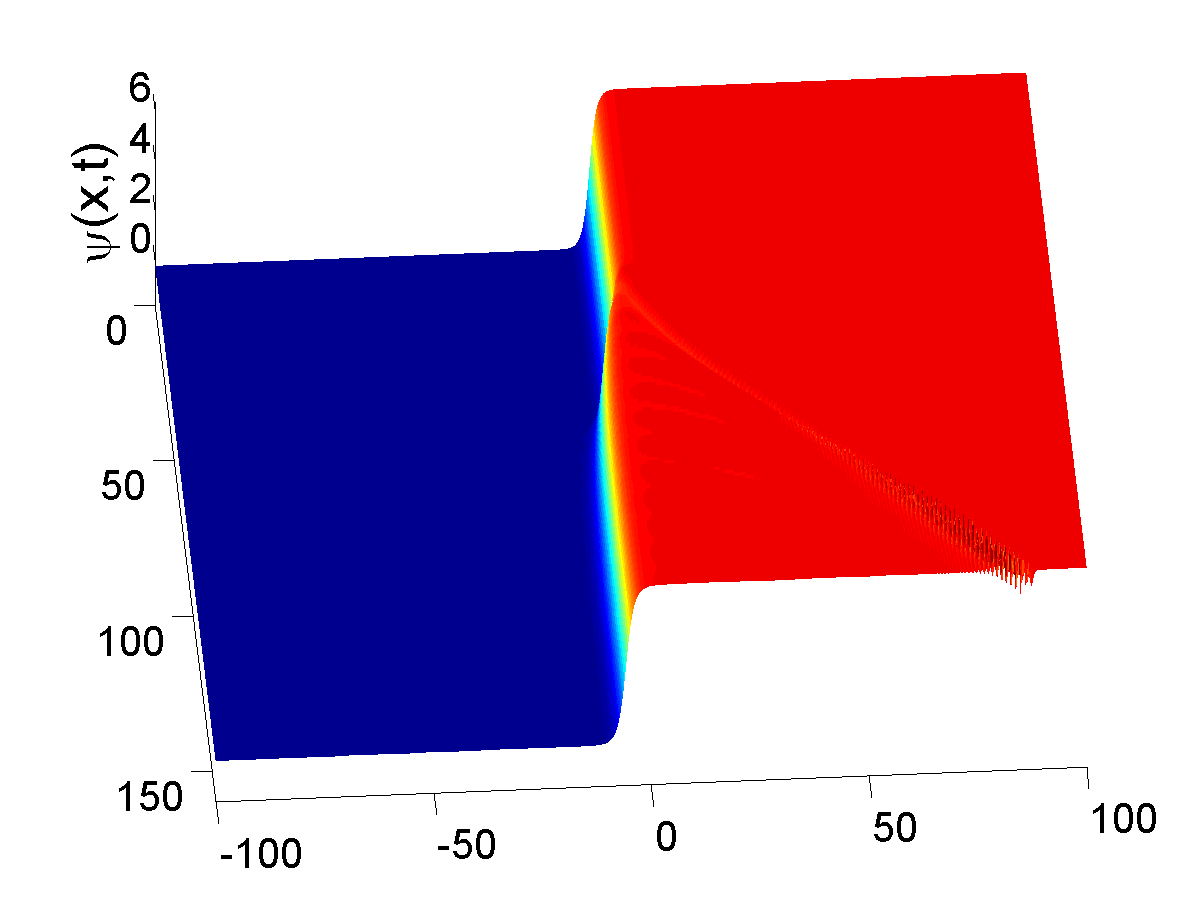} \\
&
\end{tabular}%
\caption{(Color online) The evolution of unstable KK complexes. The top and
bottom panels display components $\protect\phi (x,t)$ and $\protect\psi %
(x,t) $, respectively. In both cases, $\protect\epsilon =0.25$ and $\protect%
\beta =0.1$. In the left panels, $\protect\alpha =0$, and $\protect\alpha %
=0.05$ on the right-hand side.}
\label{fig:movingKK}
\end{figure}

\begin{figure}[tbp]
\begin{tabular}{cc}
\includegraphics[width=7cm,bb=0 0 576 432,clip=on]{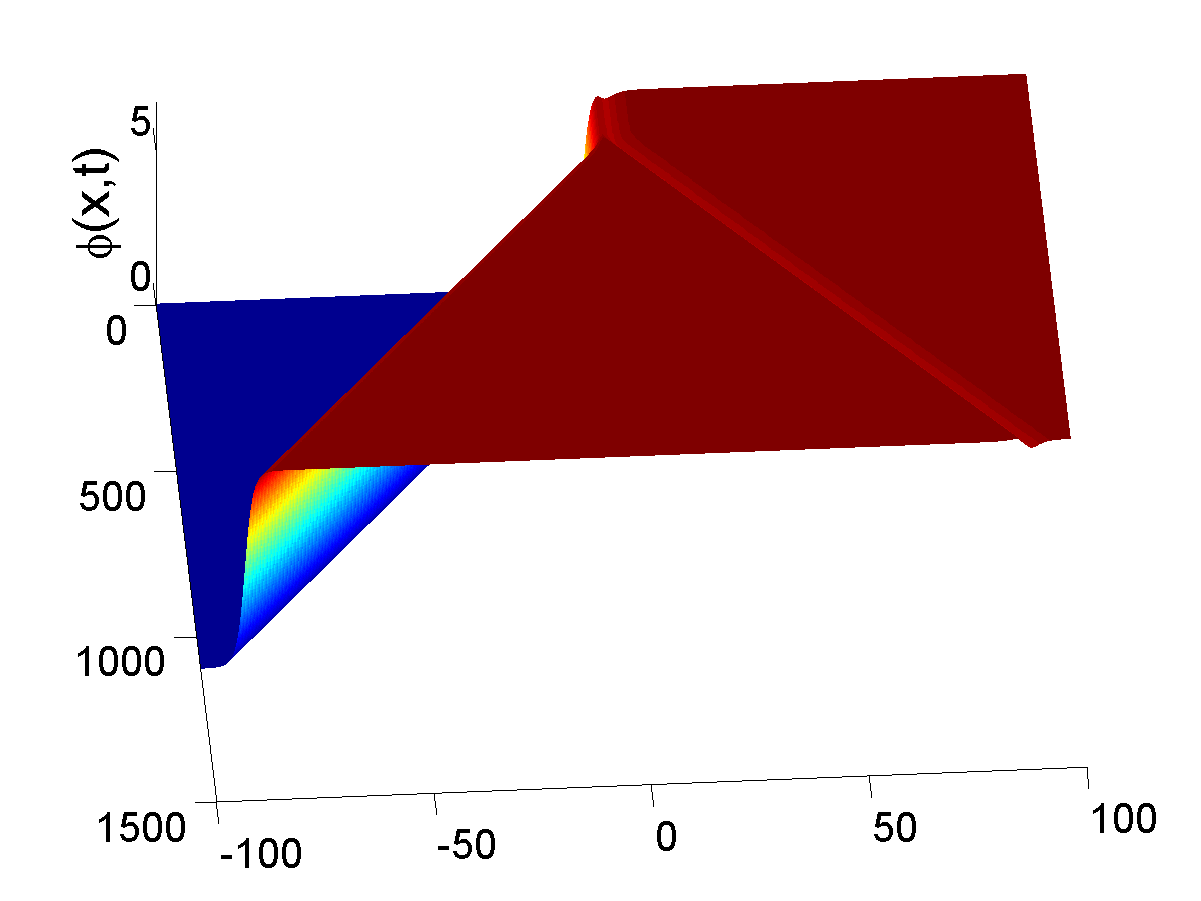} & %
\includegraphics[width=7cm,bb=0 0 576 432,clip=on]{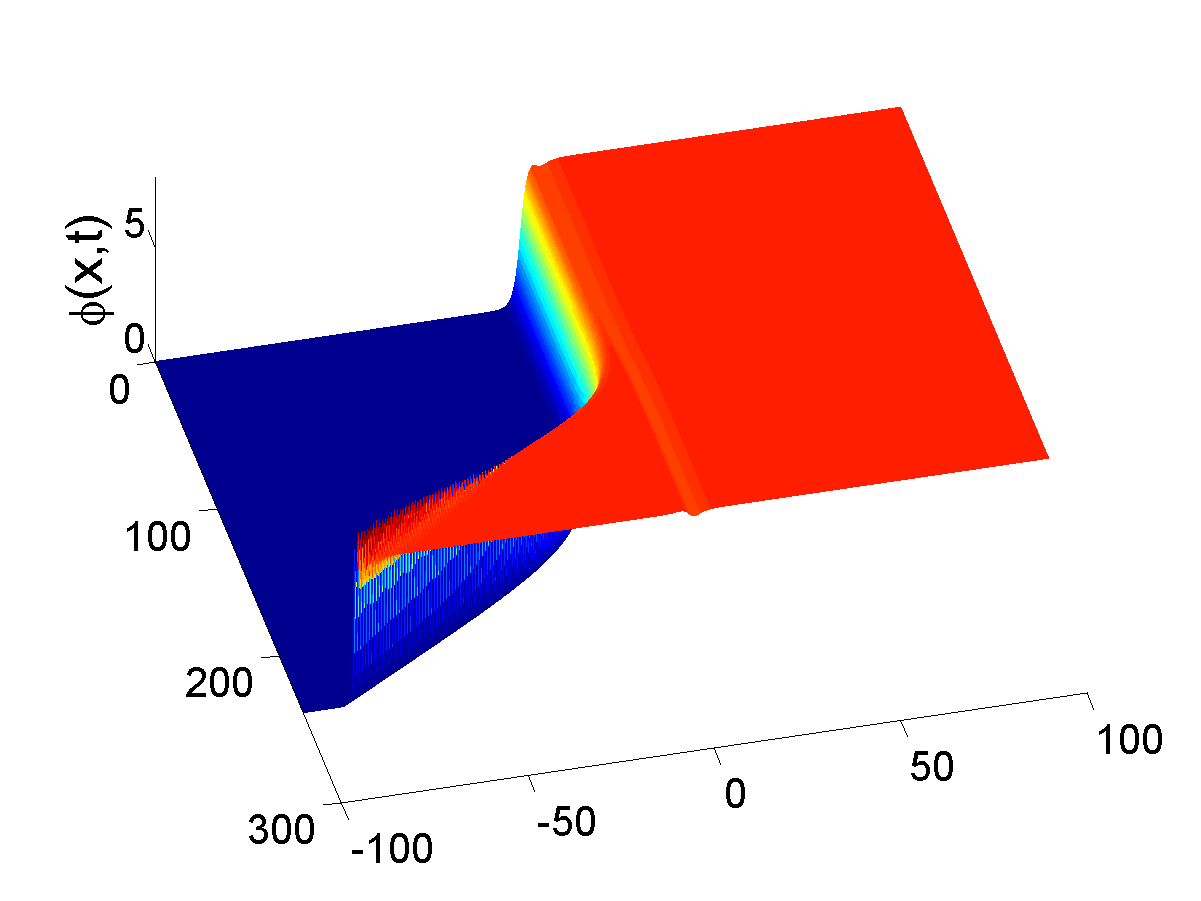} \\
\includegraphics[width=7cm,bb=0 0 576 432,clip=on]{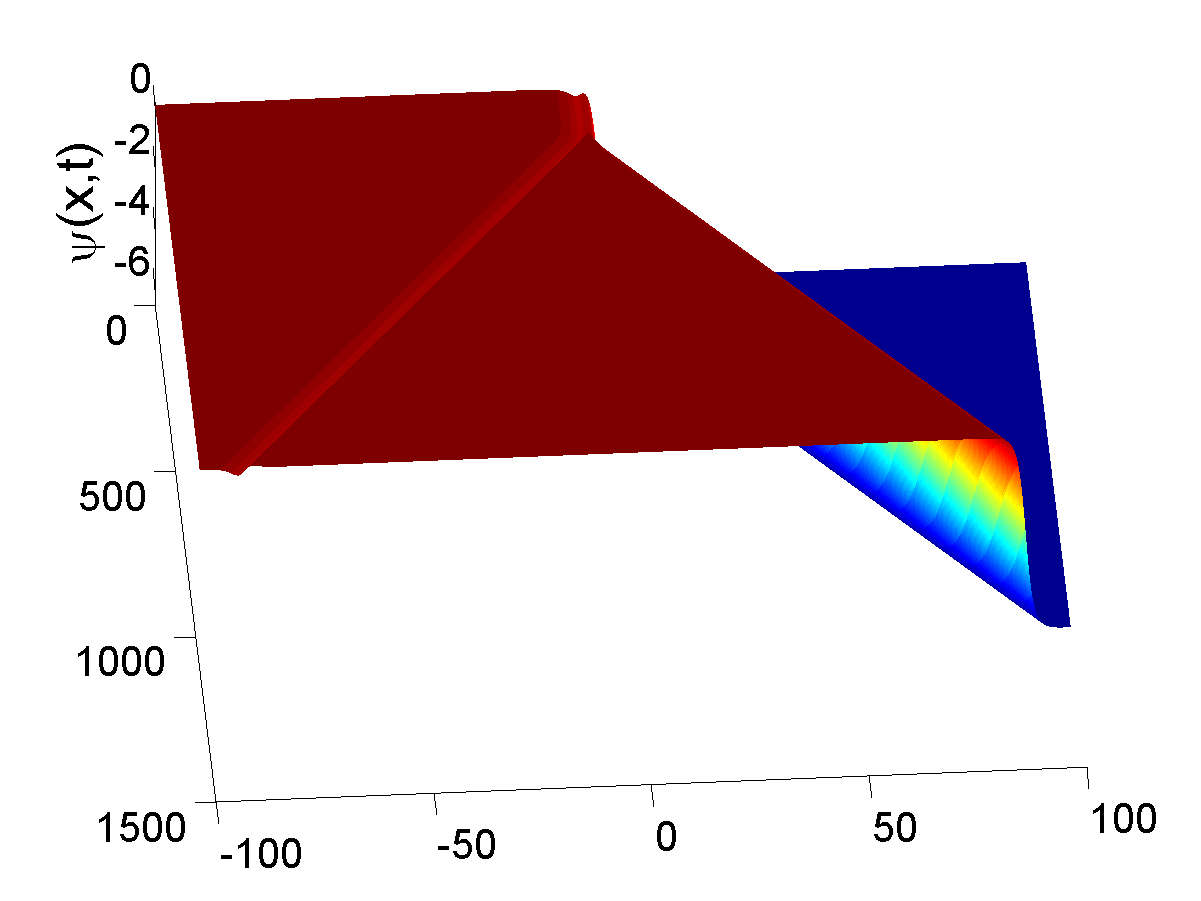} & %
\includegraphics[width=7cm,bb=0 0 576 432,clip=on]{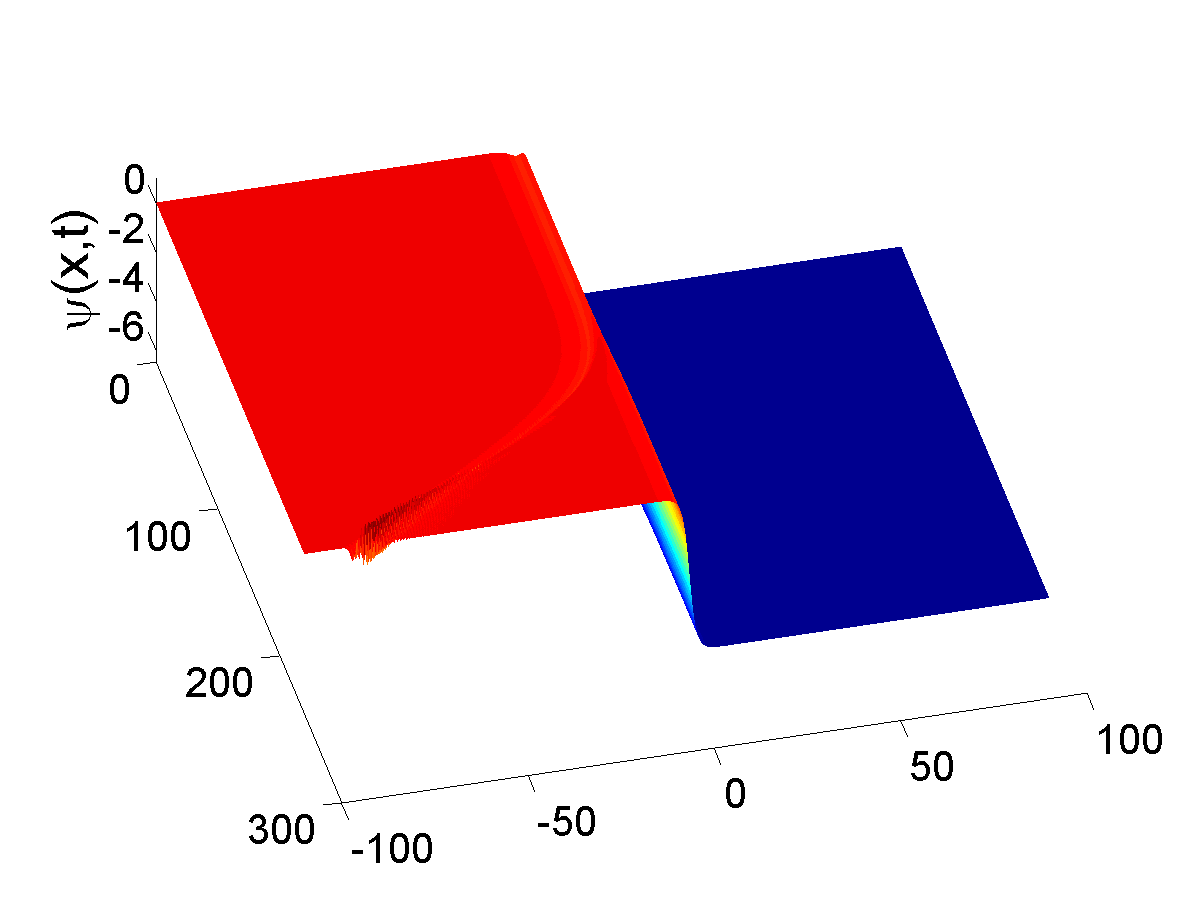} \\
&
\end{tabular}%
\caption{(Color online) The evolution of unstable KA complexes. The panels
have the same meaning as in Fig. \protect\ref{fig:movingKK}, with the same
values of $\protect\epsilon $, $\protect\beta $, and $\protect\alpha $.}
\label{fig:movingKA}
\end{figure}

This instability leads to motion of the kinks, with the different components
moving in opposite directions, i.e., the instability splits the KK and KA
complexes, as shown in Figs. \ref{fig:movingKK} and \ref{fig:movingKA}. Note
that, at $\alpha =0$, the kinks move at constant velocities, with equal
absolute values of the velocities in the two components, $\phi $ and $\psi $%
. However, at $\alpha \neq 0$, the kinks move with monotonously varying
velocities, whose absolute values in the two components are different. In
the latter case, the kink in the gain component accelerates, while that of
the lossy component --in a way reminiscent to observations in~\cite{SG-PT}--
drops its speed and finally come to a halt.

\subsection{Stable KK and KA complexes at $\protect\epsilon <0$}

\begin{figure}[tbp]
\begin{tabular}{cc}
\includegraphics[width=7cm]{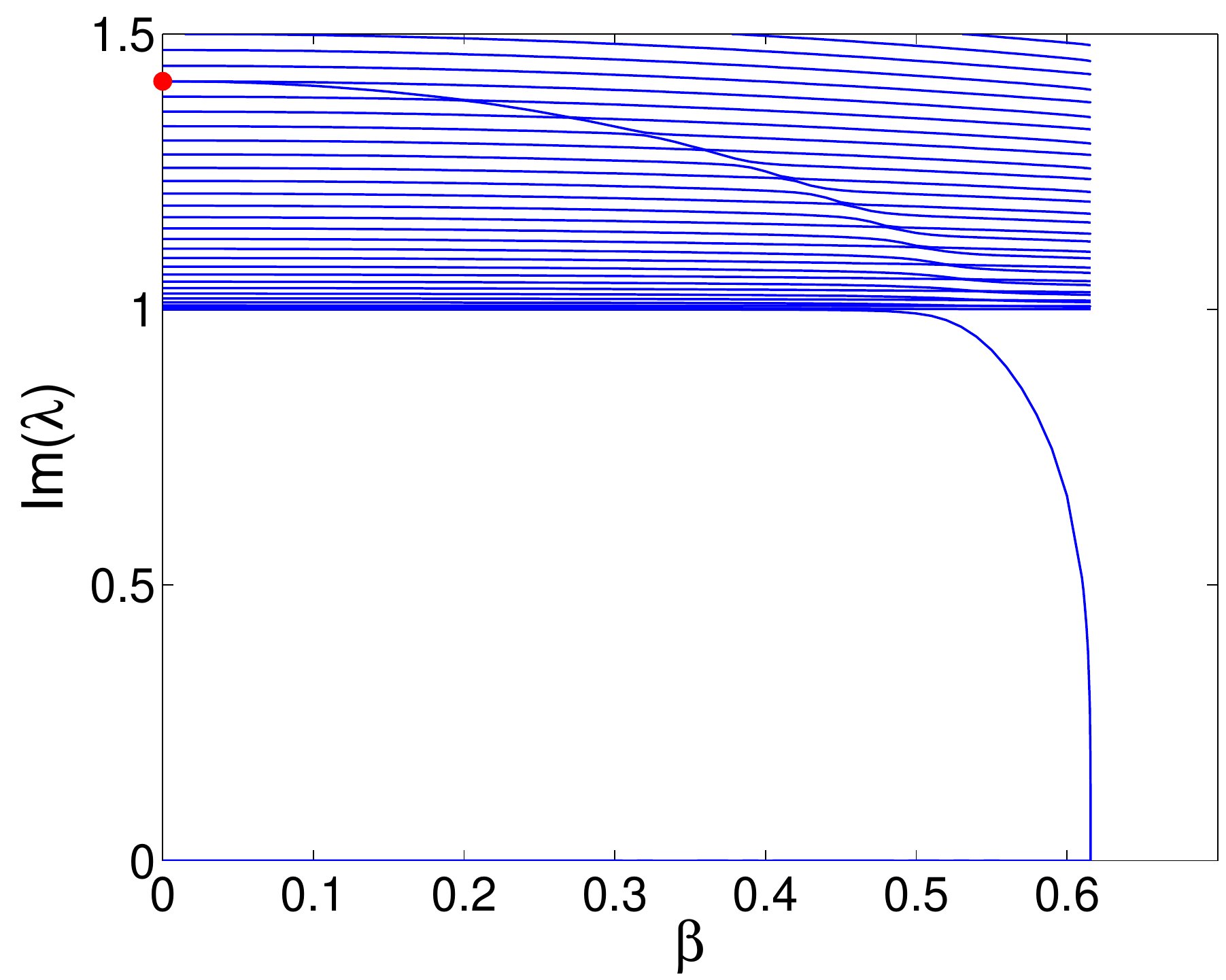} & %
\includegraphics[width=7cm]{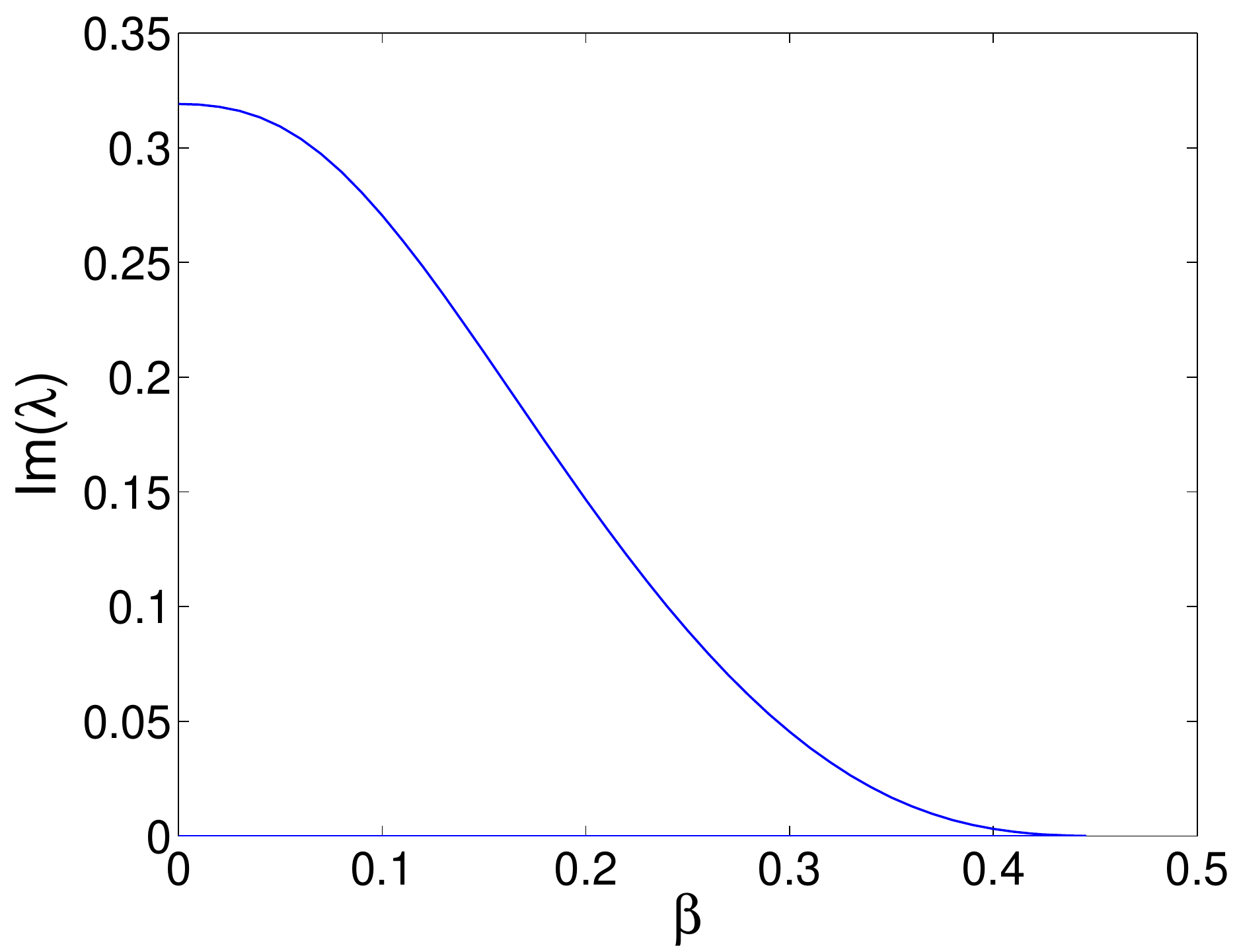}%
\end{tabular}%
\caption{(Color online) The dependence of the imaginary part of the
stability eigenvalues on $\protect\beta $ at $\protect\alpha =0$ and $%
\protect\epsilon =-1$ for \emph{stable} KK and KA complexes (left and right
panels, respectively). They exist up to the points at which Im$(\protect%
\lambda )$ vanishes. {The red dot at the left panel shows the exact
prediction for the zero mode of (\protect\ref{eps})}}
\label{fig:stab2}
\end{figure}

We have also considered the complexes in the case of $\epsilon \leq 0$.
Similar to the situation at $\epsilon >0$, bound states of both KK and KA
types exist at $\beta <\beta _{c}$. Their existence limits are also shown in
Fig. \ref{fig:plane1}, whereas Fig.~\ref{fig:stab2} showcases the dependence
of the stability eigenvalues on $\beta $ for given $\epsilon <0$ and $\alpha
=0$. It can be seen here that the imaginary eigenvalues approach $0$ as $%
\beta $ approaches $\beta _{c}$. A drastic difference from the case of $%
\epsilon >0$ is that the KK complexes are \emph{spectrally stable} whenever
they exist at $\epsilon \leq 0$, once again in the full agreement with the
analytical result for $\alpha =0$, given by Eq. (\ref{eps}). The KA modes
are \emph{stable} too if the condition $\epsilon \leq 0$ is supplemented by $%
\alpha ~\leq \alpha _{c}$, for some appropriate finite critical value of the
gain-loss coefficient, $\alpha _{c}\leq \beta $ [i.e., $\alpha _{c}$
satisfies condition (\ref{3})]. Exactly at $\epsilon =0$, the KK and KA only
exist for $\beta =0$ (notice that this is the uncoupled limit where $\phi $
and $\psi $ are independent) and they are stable only for $\alpha =0$ as, if
$\alpha \neq 0$ for $\beta =0$ the complexes cannot be stable because
conditions (\ref{3})-(\ref{4}) are violated. In this context, it is perhaps
more precise to say that one component is subject to damping, while the
other subject to pumping, with the latter featuring unstable dynamics.

\begin{figure}[tbp]
\includegraphics[width=7cm]{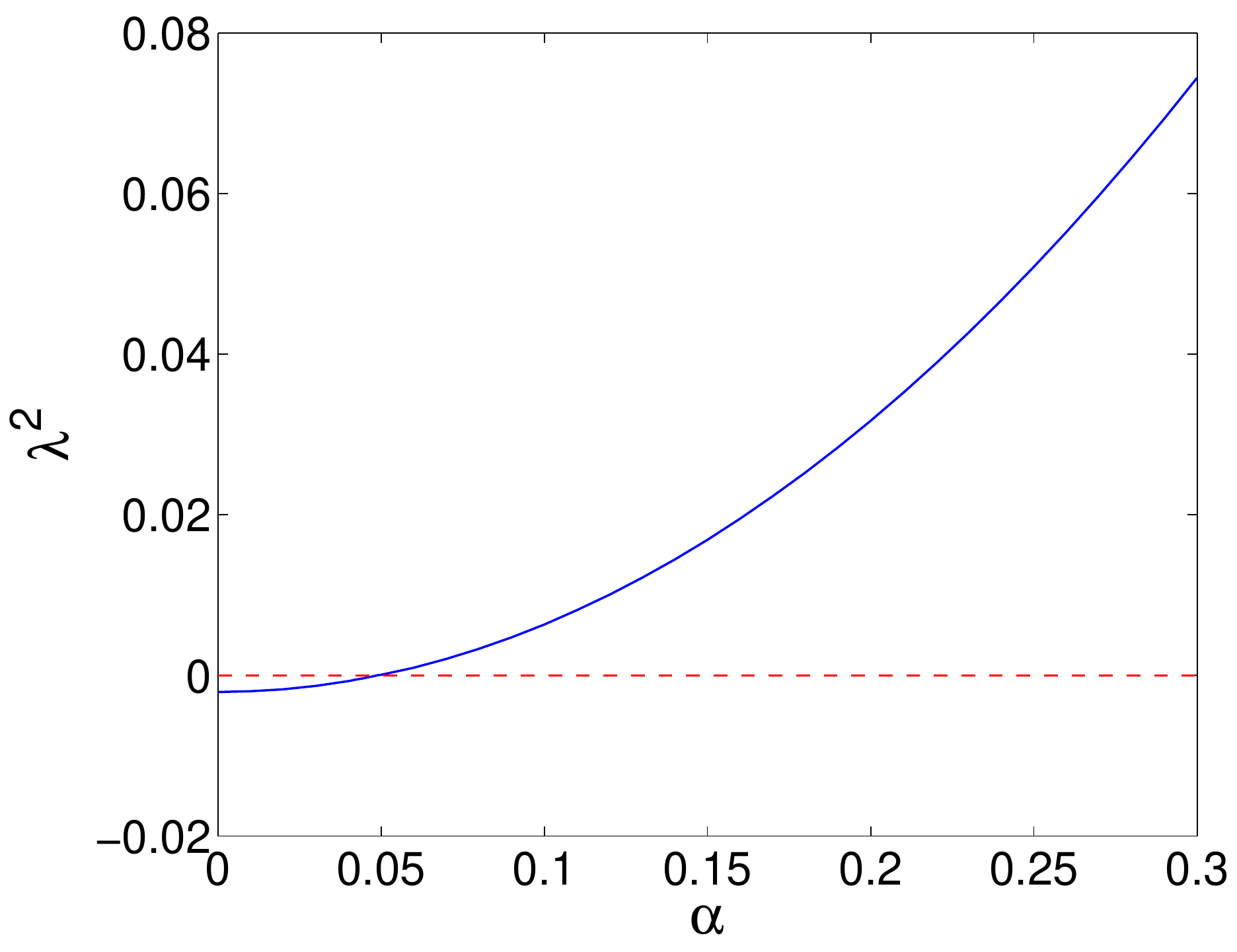}
\caption{(Color online) {Squared instability eigenvalue for the KA
complexes, versus the gain-loss coefficient, $\protect\alpha $, at $\protect%
\beta =0.3$ and $\protect\epsilon =-1$.}}
\label{fig:stab3}
\end{figure}

\begin{figure}[tbp]
\begin{tabular}{cc}
\includegraphics[width=7cm,bb=0 0 576 432,clip=on]{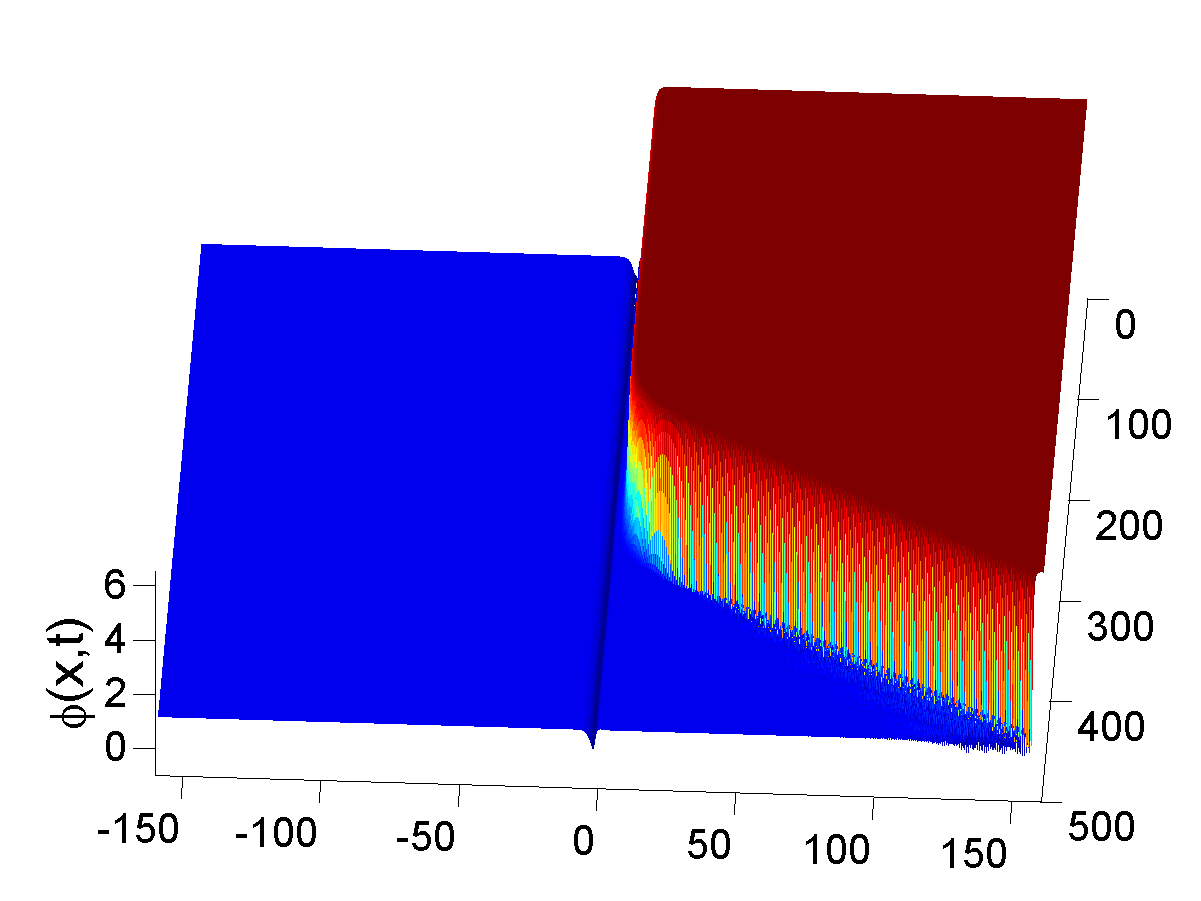} & %
\includegraphics[width=7cm,bb=0 0 576 432,clip=on]{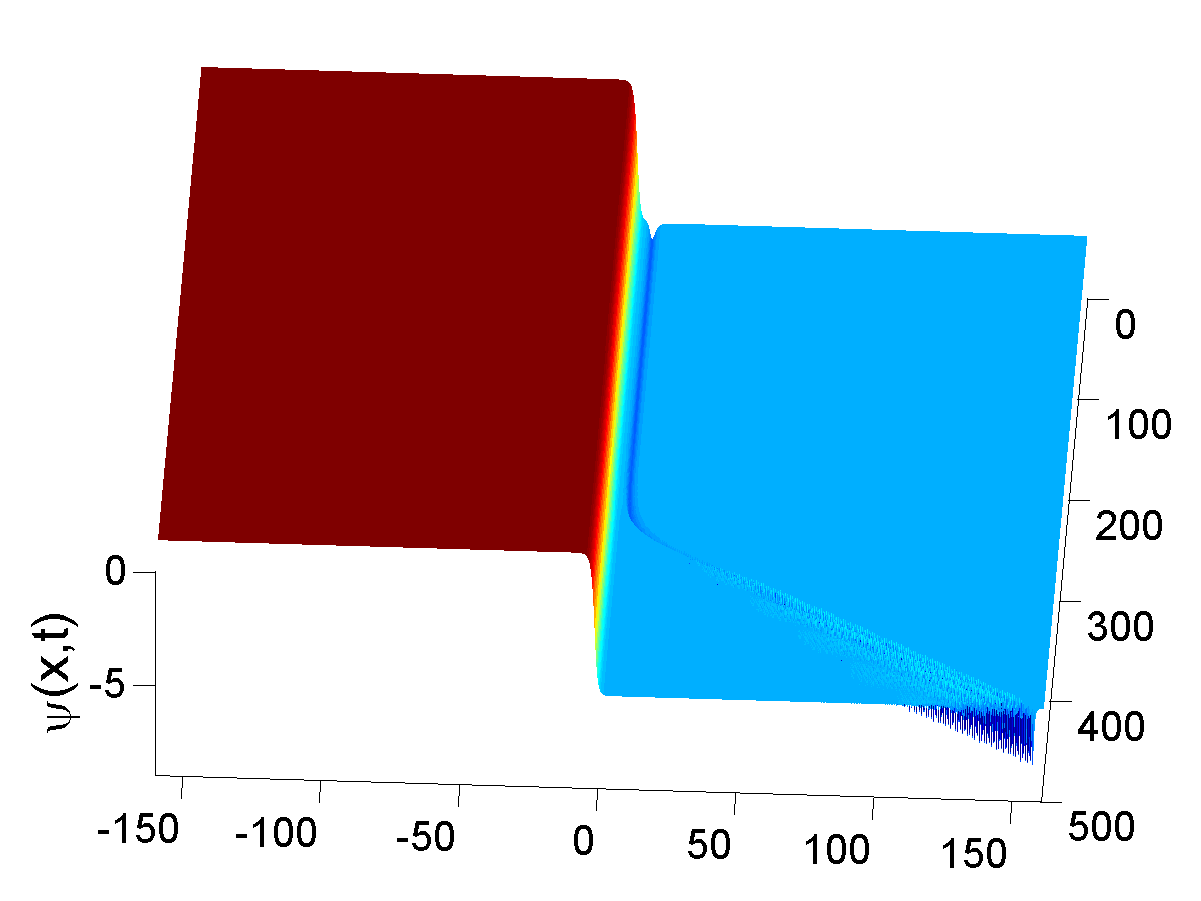} \\
&
\end{tabular}%
\caption{(Color online) The evolution of an unstable KA solution at $\protect%
\epsilon =-1$, $\protect\beta =0.3$ and $\protect\alpha =0.08$. To amplify
effects of the true exponential instability, against the spurious
instability of the background (see the text), the onset of the instability
was catalyzed by adding an initial perturbation proportional to the
corresponding eigenmode.}
\label{fig:movingneg}
\end{figure}

The KA solutions become exponentially unstable at $\alpha >\alpha _{c}$, as
shown in Figs. \ref{fig:stab3} and \ref{fig:movingneg}. The latter figure
shows that the instability splits the KA complex into two components.
Naturally, the kink in the component ($\psi $) which is subject to the
action of the dissipation comes to a halt, while its counterpart in the
amplified component ($\phi $) becomes a traveling one, accelerating over
time. Each of the separated kinks creates its \textquotedblleft shadow" in
the mate component, in the form of a dip.

The dependence of $\alpha _{c}$ on $\beta $ is presented in Fig. \ref%
{fig:plane}. A noteworthy feature of this dependence is its non-monotonous
form, with a maximum of $\alpha _{c}$ attained at a particular value of $%
\beta $, which depends on $\epsilon $. This maximum is caused by the fact
that, on the one hand, if $\beta $ falls below a critical value, then $%
\alpha _{c}>\beta $ and condition (\ref{3}) does not hold, hence the
background (flat-state's) instability masks the exponential instability; on
the other hand, if $\beta$ is above that critical value, then $\beta >\alpha
_{c}$ and the latter instability manifests itself.

\begin{figure}[tbp]
\includegraphics[width=7cm]{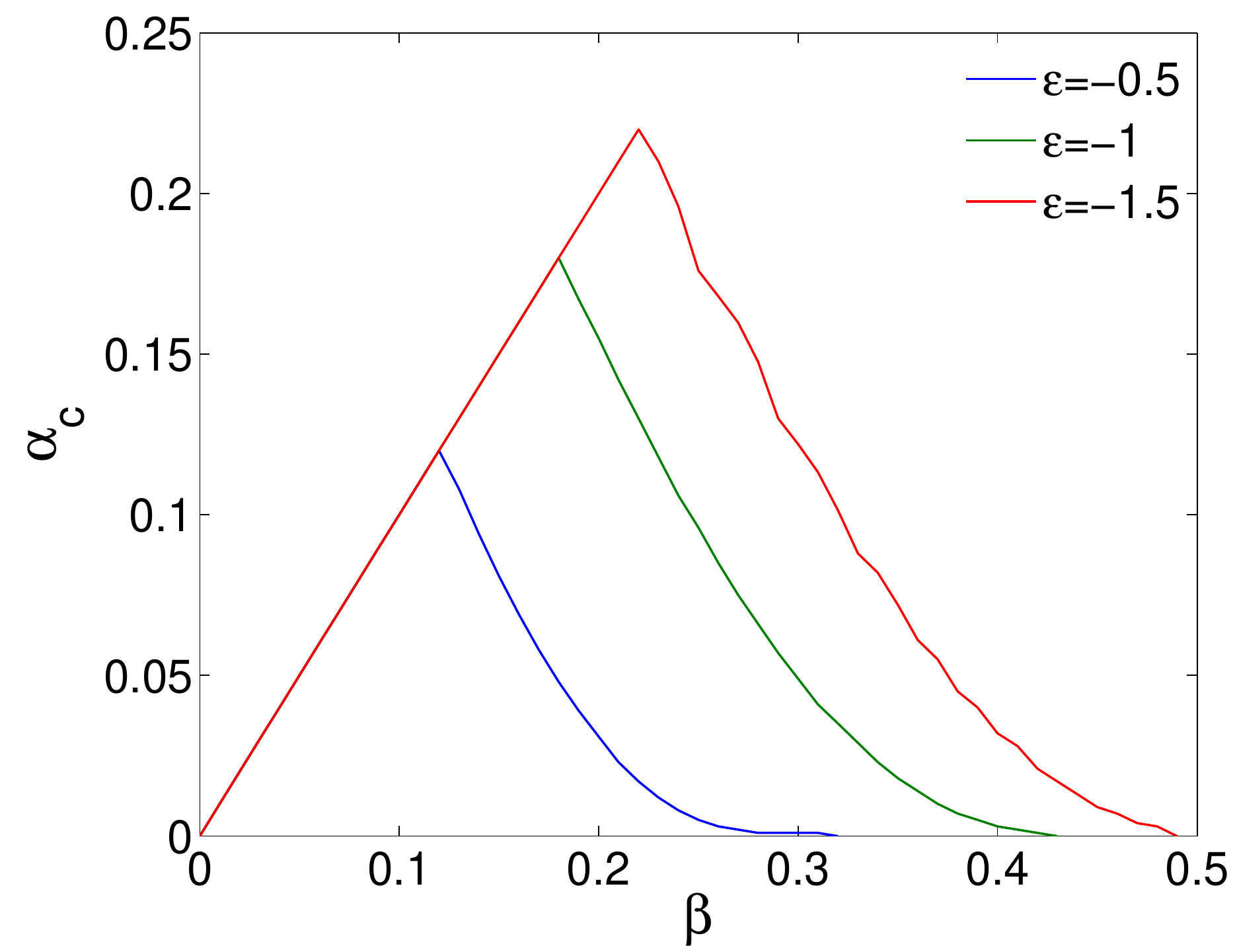}
\caption{(Color online) The critical value of the gain-loss coefficient, $%
\protect\alpha _{c}$, above which the KA complexes are unstable at $\protect%
\epsilon <0$, versus $\protect\beta $.}
\label{fig:plane}
\end{figure}

From a technical standpoint, it is relevant to note in passing that the
finite-difference discretization of the first-order spatial and temporal
derivatives introduces a number of additional, yet spurious numerical
instabilities (associated with complex eigenvalues stemming from the
continuous spectrum), disappearing as one approaches the continuum,
infinite-domain limit; see~Ref. \cite{SG-PT} for similar examples with
temporal first derivatives in $\mathcal{PT}$-symmetric sine-Gordon and
related systems, and~Ref. \cite{Dirac} for such examples with spatial first
derivatives in problems featuring Dirac operators. In what we have discussed
above, we have not considered these instabilities, focusing on the true
dynamical features of the continuum problem.

\section{Conclusions}

We have introduced a system of coupled sine-Gordon equations, with mutually
balanced gain and loss in them, which represents the $\mathcal{PT}$ symmetry
in the system. The consideration of this system helps to understand
possibilities for the implementation of the $\mathcal{PT}$ symmetry, which
was elaborated in detail in optics, in other physical settings. Two types of
coupling were included: sinusoidal terms, and the cross-derivative coupling.
The former coupling corresponds to a commonly utilized interaction term
between the corresponding FK (Frenkel-Kontorova)\ chains. The
cross-derivative coupling was not considered in previously studied models.
It may be generated by three-body interactions, assuming that the particles
belonging to parallel chains move along different directions. The $2\pi $ KK
(kink-antikink) and KA (kink-antikink) complexes were constructed in the
system, and their stability was studied, by means of analytical and
numerical methods. It has been found that the complexes are stable or
unstable, depending on the sign of the sinusoidal coupling term. Stability
regions for the complexes of both types were identified in the parameter
space of the temporal gain/loss and the cross-derivative coupling strength.
Simulations reveal splitting of unstable complexes into separating kinks and
antikinks. The latter move at constant speed in the absence of gain and
loss, and follow the dynamics imposed by the gain (acceleration) or loss
(deceleration) in their presence.

There remain numerous interesting issues to address as a continuation of the
present work. In particular, the possibility of the existence of stable
traveling KK and KA complexes is a challenging problem. Moreover, a
potentially analytical or semi-analytical study in the spirit of~Ref. \cite%
{galley} could provide a set of guidelines on the expected motion (and
stability) of the kinks in the presence of the additional terms considered
herein. It is interesting too to identify stability boundaries for $\mathcal{%
PT}$-symmetric solitons of the NLS type in the framework of Eqs. (\ref{u})
and (\ref{v}) with $\beta \neq 0$.

\authorcontributions{B.A.M. proposed the model and performed analytical considerations. J.C-M. carried
out the numerical work. P.G.K. contributed to all parts of the work. All the
authors have equally contributed to drafting the paper.}

\conflictofinterests{The authors declare no conflict of interest.}

\renewcommand\bibname{References}

\end{document}